\documentclass{WileyMSP-template}
\usepackage{amssymb}
\usepackage{amsmath}
\usepackage{comment}
\usepackage[thinc]{esdiff}
\usepackage{amsmath}
\usepackage{bm}
\usepackage{amsfonts}
\usepackage{graphicx}
\usepackage{subcaption}
\usepackage{xcolor}
\usepackage{hyperref}
\usepackage{academicons}

\usepackage{multirow}
\usepackage{float}
\usepackage{array}

\usepackage{hyperref}
\usepackage{academicons}
\usepackage{xcolor}

\usepackage{hyperref}
\hypersetup{
    colorlinks=true,
    linkcolor=blue,
    filecolor=magenta,      
    urlcolor=cyan,
    pdftitle={ORCID Integration},
    bookmarks=true,
    pdfpagemode=FullScreen,
}

\usepackage{xcolor}

\newcommand{\orcid}[1]{\href{https://orcid.org/#1}{\raisebox{-0.45ex}{\includegraphics[width=1.8ex]{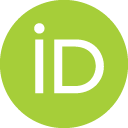}}}}

\begin{document}

\pagestyle{fancy}
\rhead{\includegraphics[width=2.5cm]{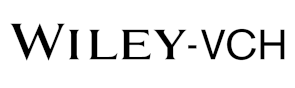}}

\title{Leveraging Transfer Learning to Overcome Data Limitations in Czochralski Crystal Growth}

\maketitle

\author{Milena Petkovic*\orcid{0000-0003-1632-4846}}
\author{Natasha Dropka\orcid{0000-0002-1530-7449}}
\author{Xia Tang}
\author{Janina Zittel\orcid{0000-0002-0731-0314}}

% Affiliations: Make sure each corresponding author has their email provided
\begin{affiliations}
Milena Petkovic*, Natasha Dropka and Xia Tang\\
Leibniz-Institut für Kristallzüchtung, Max-Born Str. 2, Berlin, Germany\\
Email: milena.petkovic@ikz-berlin.de\\
Janina Zittel\\
Zuse Institute Berlin, Takustraße 7, Berlin, Germany
\end{affiliations}
% Keywords: Please provide a minimum of three and a maximum of seven keywords, separated by commas


% Abstract should be written in the present tense and impersonal style (i.e., avoid we), and be at most 200 words long

\begin{abstract}

The Czochralski (Cz) method is a widely used process for growing high-quality single crystals, critical for applications in semiconductors, optics, and advanced materials. Achieving optimal growth conditions requires precise control of process and furnace design parameters. Still, data scarcity -- especially for new materials -- limits the application of machine learning (ML) in predictive modeling and optimization. This study proposes a transfer learning approach to overcome this limitation by adapting ML models trained on a higher data volume of one source material (Si) to a lower data volume of another target material (Ge and GaAs). The materials were deliberately selected to assess the robustness of the transfer learning approach in handling varying data similarity, with Cz-Ge being similar to Cz-Si, and GaAs grown via the liquid encapsulated Czochralski method (LEC), which differs from Cz-Si. We explore various transfer learning strategies, including Warm Start, Merged Training, and Hyperparameters Transfer, and evaluate multiple ML architectures across two different materials. Our results demonstrate that transfer learning significantly enhances predictive accuracy with minimal data, providing a practical framework for optimizing Cz growth parameters across diverse materials.

\end{abstract}



\section{Introduction}
\label{Sec.Introduction}

For over a century, the Czochralski (Cz) method has been used to produce large, high-quality single crystals, which have been crucial for modern electronic and photonic applications~\cite{Rudolph2014, Muller2007}. The quality and yield of these crystals are significantly influenced by factors such as furnace design -- including the geometry and material properties of hot-zone components -- and process parameters like pulling rate, rotational rates, and heating power. Extensive experimental and numerical studies have been conducted to optimize these variables, particularly in silicon Czochralski growth (Cz-Si), e.g.~\cite{Noghabi2013, Ding2018, Mosel2020}, given its prominence in the semiconductor industry. Parallel research efforts have also been directed toward other materials, various
semiconductors, binary compound semiconductors, oxides, and fluorides, to enhance their crystal growth processes, e.g.~\cite{Friedrich2015, DEPUYDT2006437, Sarukura2015, Bottcher1999}.

In recent years, machine learning (ML) has emerged as a powerful tool for modeling, monitoring, and optimizing bulk crystal growth processes. Data-driven ML models offer predictive capabilities that complement traditional physics-based approaches such as  Computational Fluid Dynamics (CFD). However, insufficient or incomplete data often restricts ML application in crystal growth, especially for new materials. Training robust ML models typically requires large, high-quality datasets covering a wide range of processing conditions. In practice, the data is often unavailable due to the high costs of crystal growth experiments and their time-consuming nature.

Studies have increasingly incorporated ML techniques like artificial neural networks (ANNs), genetic algorithms (GAs), decision trees, and long short-term memory (LSTM) models. For instance, Qi et al.~\cite{QI2020} used ANN and GA to optimize melt/crystal interface shape and oxygen concentration in Cz-Si. Ding and Liu~\cite{Ding2018} applied ML for real-time prediction of the interface shape to adjust process parameters dynamically. Kutsukake et al.~\cite{Kutsukake2020} further extended this approach by using ML for the real-time prediction of interstitial oxygen concentration in Czochralski-Silicon (Cz-Si) growth, helping to enhance control over oxygen-related defects.
Research by Dropka et al.~\cite{Dropka2021a} combined decision trees with data mining techniques to optimize GaAs crystal growth, illustrating how ML can derive critical insights from small datasets. 
Yu et al.~\cite{Yu2021} demonstrated how ML can guide furnace design optimization, using ANN and genetic algorithm to optimize the geometrical design of a crystal growth system, which led to improved production efficiency. These advances complement more traditional CFD approaches, as illustrated in Cz-Ge furnace optimization studies by Dropka et al.~\cite{Dropka2022}, which blend ML methods with experimental designs to understand the relationships between furnace geometry, process parameters, and crystal quality.

The fusion of machine learning (ML) with computational fluid dynamics (CFD) has opened new avenues for optimizing Czochralski crystal growth processes. While CFD simulations have traditionally been employed to model the complex transport phenomena during crystal growth, they often demand substantial computational resources and time. Moreover, these simulations can be prohibitively expensive and usually struggle to provide generalized solutions applicable to a wide range of scenarios. To accelerate data generation, the complexity of the CFD model is often reduced by neglecting melt convection and/or considering only a local domain of the melt and crystal in the model~\cite{Vieira2025}. This limitation highlights the need for more efficient and adaptable approaches to studying crystal growth processes. 

We propose a transfer learning approach to tackle this challenge to improve ML predictions in Cz crystal growth. Transfer learning enables ML models to leverage knowledge from a well-studied source material and apply it to a target material with limited data, reducing the need for extensive new training datasets. Although transfer learning has been widely explored in natural language processing and computer vision, its application in bulk crystal growth and process modeling remains relatively unexplored. In the study conducted by Dang et al.~\cite{Dang2022}, the authors employed convolutional neural networks (CNNs) for transfer learning to predict the geometric evolution of silicon carbide (SiC) single crystals during an unsteady solution growth process, significantly improving the prediction accuracy of crystal morphology while minimizing the reliance on extensive experimental datasets. 

\begin{figure}
\centering
  \includegraphics[width=0.6\linewidth]{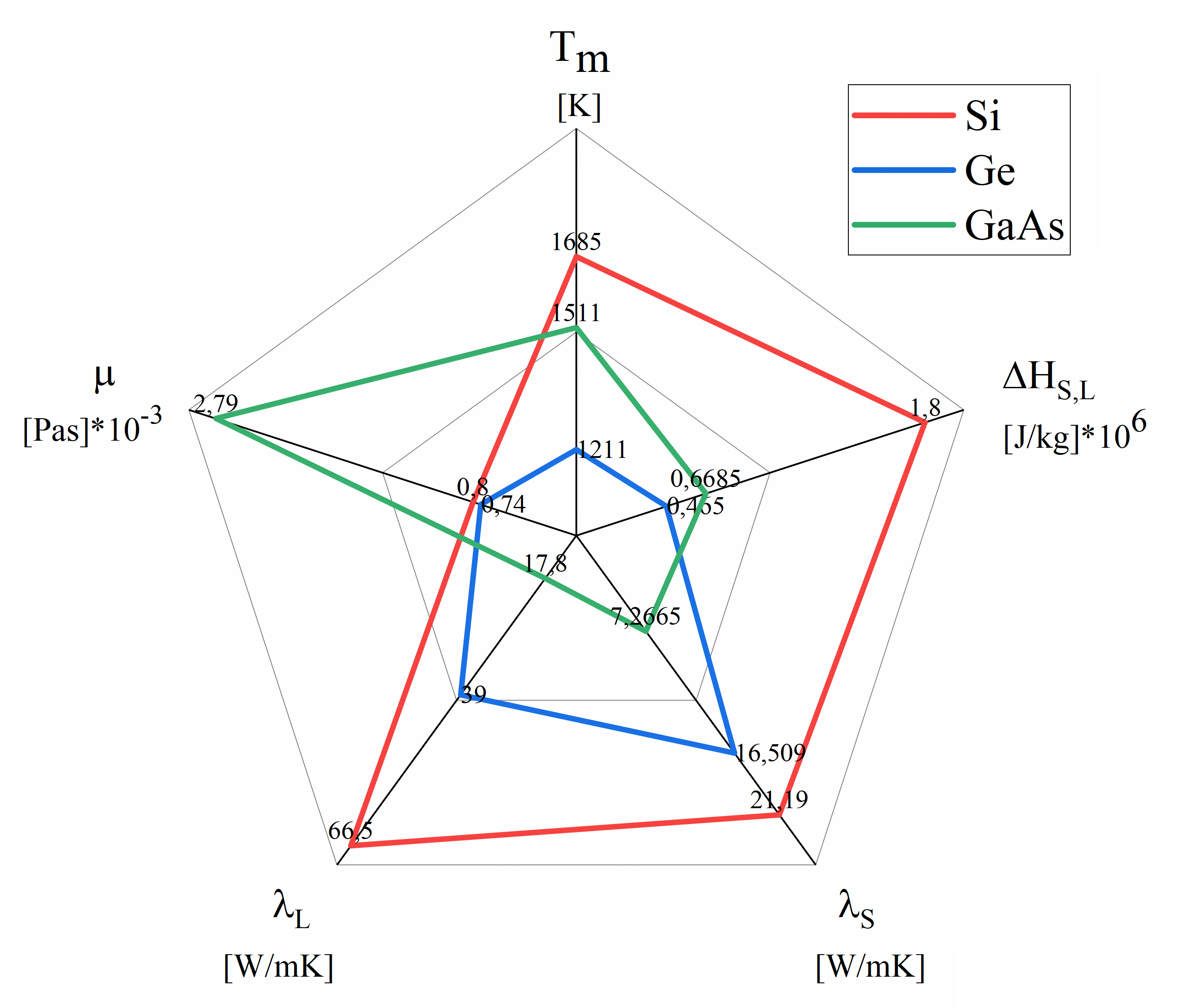}
  \caption{ Key material properties of Si, Ge, and GaAs: melting point $T_m$, latent heat of solidification $\Delta H_{S,L}$, dynamic viscosity $\mu$, and heat conductivity of crystal $\lambda_S$ and melt $\lambda_L$.}
  \label{fig:materialproperties}
\end{figure}

In this study, we focus on three technologically important crystalline materials: Silicon (Si), Germanium (Ge), and Gallium Arsenide (GaAs). 
These materials were deliberately selected to assess the robustness of the transfer learning approach in handling varying data similarity, both in terms of the growth process parameters and the material properties of the grown material. Figure~\ref{fig:materialproperties} shows the rotor plot of key material properties for Si, Ge, and GaAs. Regarding thermal properties, Si and Ge are more similar than Si and GaAs~\cite{Frank-Rotsch2018}. This similarity stems from their shared diamond cubic crystal structure, which results in comparable thermal conductivities, thermal expansion behaviors, and stability at high temperatures.

Regarding the process, Cz-Ge is similar to Cz-Si, while GaAs, grown via the LEC method~\cite{Mullin2015}, differs from Cz-Si. The LEC method is a variation of Cz that involves encapsulating the melt with a protective layer of a high-melting-point material. LEC is often used for compound semiconductors because the encapsulation helps maintain the proper composition and minimizes loss due to vaporization. The growth rate in the LEC method is typically slower because of the additional complexity of the encapsulation layer and the more controlled growth conditions.

The proposed data-driven approach may be applied to different types of datasets, taking into account their fidelity and origin (e.g., a high number of synthetic data versus a low number of experimental data) and the material being grown (e.g., abundant data points for one material versus scarce data for another, whether synthetic or experimental). This will be demonstrated in the present study. Given the extensive data available for Si compared to Ge and GaAs, we explore how transfer learning can be utilized to adapt ML models trained on Si to predict and optimize the growth processes of Ge and GaAs.

To overcome the limitations of scarce experimental data, we generate training data using computational fluid dynamics (CFD) simulations of the Cz process. CFD provides detailed, physics-based insights into crystal growth processes, allowing the generation of synthetic datasets covering a wide range of process conditions without the constraints of physical experimentation. By leveraging CFD-generated data, we explore several transfer learning strategies, including warm start, merged training, and hyperparameters transfer, and evaluate multiple ML architectures.

The remainder of this paper is organized as follows: Section~\ref{sec:methodology} describes our dataset and methodology, detailing the CFD simulations, ML models, and transfer strategies employed. Section~\ref{sec:results} presents our simulation and ML results and give comparative analysis. Finally, Section~\ref{sec:conclusion} concludes with key insights and future research directions.

\section{Methodology}
\label{sec:methodology}
\subsection{CFD numerical model for data generation}\label{Sec.CFDnumericalmodel}

\begin{figure}[H]
\centering
  \includegraphics[width=0.4\linewidth]{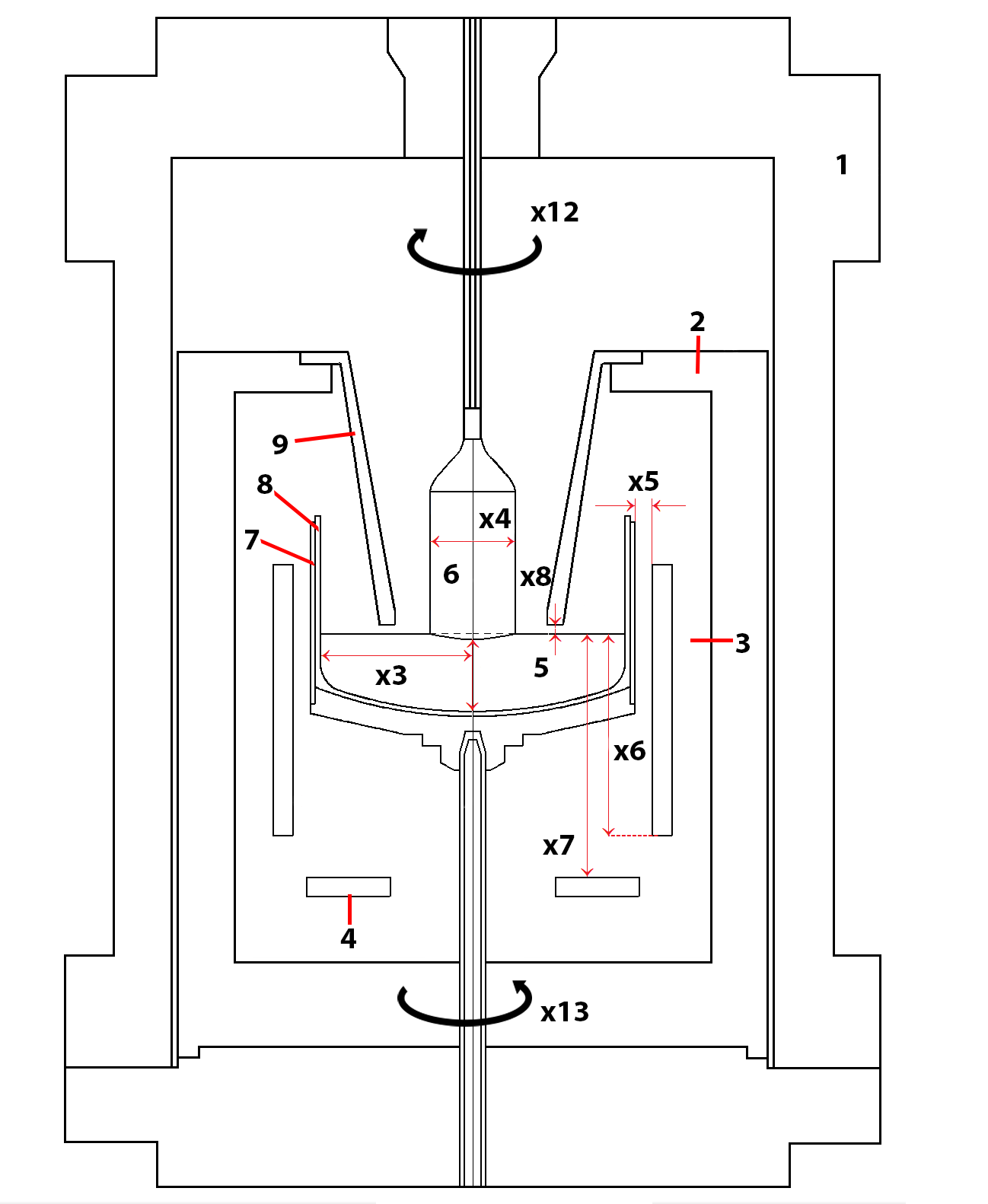}
  \caption{Schematic view of Cz setup. 1-steel casing, 2-graphite based insulation, 3-graphite side heater, 4-graphite bottom heater, 5- melt, 6-crystal, 7-graphite crucible support, 8-quartz crucible and 9-radiation shield made of various materials. Parameters $x_i$ are marked in the picture and described in Table~\ref{Table:parameters}
}
  \label{fig:furnance}
\end{figure}

Numerical axisymmetric quasi-steady-state CFD simulations has been conducted to model Cz crystal growth in a furnace equipped with a bottom and a side resistance graphite heaters, as shown in Figure~\ref{fig:furnance}. The furnace also featured a radiation shield made from one of five materials: graphite,
quartz, SiC, TiC, or ceramic, along with a crucible composed of quartz, graphite, or BN, depending on
the material being grown. Carbon-based insulation and an Ar gas atmosphere were
used for all materials. In the case of LEC-GaAs growth, a B$_2$O$_3$ encapsulant was employed. The material properties
utilized in this study originated from~\cite{Dropka2024,Petkovic2025}.
The CFD model for all grown materials included several physical processes, such as buoyancy-driven
convection, forced convection due to crystal and crucible rotation, phase transitions, thermal
radiation, and heat transfer between the fluid and solid phases.

For each crystalline material grown by Cz-method, unique growth scenarios were simulated, varying 15 parameters listed in Table~\ref{Table:parameters} alongside the input parameter ranges for all three grown materials.

\begin{figure}[H]
\centering
  \includegraphics[width=0.35\linewidth]{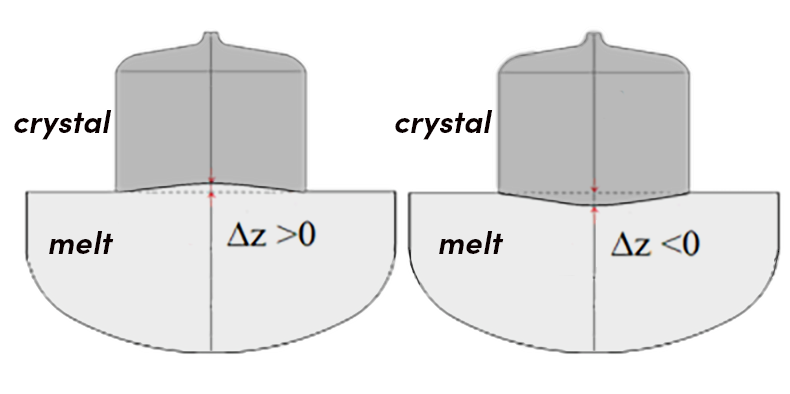}
  \caption{Concave and convex interface shapes measured by the interface deflection}
  \label{fig:InterfaceShape}
\end{figure}

The results from the CFD simulations were presented as the central interface deflection ($\Delta z$), and the Voronkov ratio $v/G$ ~\cite{Voronkov1982, Voronkov1999} along the symmetry axis. Interface deflection was measured relative to the three-phase junction and varied from concave ( $\Delta z <0 $) to convex ($\Delta z >0 $) shapes, as illustrated in Figure~\ref{fig:InterfaceShape}.

\begin{table}[ht!]
\centering
\scalebox{0.9}{

\begin{tabular}{llllll}
 & \textbf{Furnace design parameters}   & \textbf{Unit} & \multicolumn{3}{c}{\textbf{Range}}             \\ \hline
\textbf{}    & \textbf{}                            & \textbf{}     & \textbf{Si}     & \textbf{Ge}  & \textbf{GaAs} \\ \hline
$x_1$        & Material weight                      & kg            & 2.26 - 68.15    & 3.78 - 17.39 & 3.11 - 16.65  \\
$x_2$        & Fraction of material crystallized     & -             & 0.02 - 0.71     & 0.03 - 0.64  & 0.04 - 0.53   \\
$x_3$        & Crucible diameter                    & inch          & 6.32 - 16.28    & 4.36 - 9.51  & 4.36 - 9.51   \\
$x_4$        & Crystal diameter                     & inch          & 1.37 -11.87     & 1.21 - 4     & 1.21 - 4      \\
$x_5$        & Radial distance between the crucible support and side heater        & mm            & 10 - 60         & 10-60        & 10-60         \\
$x_6$        & Axial displacement of the side heater from melt free surface       & mm            & 50.16 - 180     & 92.11 - 238  & 37.11 - 238   \\
$x_7$        & Axial displacement of the bottom heater from melt free surface      & mm            & 78.26 - 208     & 101.26 - 220 & 92.78 -220    \\
$x_8$        & Axial displacement of the radiation shield from melt free surface          & mm            & 2.5 -65.5       & 8.53 - 63.53 & 11.07 - 68.53 \\
$x_9$        & Radiation shield material emissivity & -             & 0.3 -0.9        & 0.3 - 0.85   & 0.30 - 0.85   \\
$x_{10}$     & Lambda radiation shield               & W/mK          & 2.15 - 38.77    & 2.15 - 56.12 & 2.15 - 56.12  \\ \hline
             & \textbf{Process parameters}          &               &                 &              &               \\ \hline
$x_{11}$     & Pulling rate                         & mm/min        & 0.001 - 2       & 0.05 - 1.17  & 1 - 15        \\
$x_{12}$     & Crystal rotation rate                & rpm           & 1 - 60          & 1 - 60       & 5 - 35        \\
$x_{13}$     & Crucible rotation rate               & rpm           & -20 - $10^{-5}$ & -30 - (-1)   & -20 - (-1)    \\
$x_{14}$     & Side heater power                    & kW            & 0 - 58          & 0.50 - 5     & 1 - 22        \\
$x_{15}$     & Bottom heater power                  & kW            & 6.92 - 89       & 0.11 - 4.07  & 3.51 - 38.67  \\ \hline
\end{tabular}
}

\caption{Furnace design and process parameters for CFD data generation}
\label{Table:parameters}
\end{table}

The growth transport phenomena were governed by continuity equations, Navier-Stokes equations with the Boussinesq approximation, and energy balance equations. For detailed descriptions of these equations, the reader is referred to~\cite{Kakimoto2015}.

Equations~\ref{eqn:1}--\ref{eqn:3} describe the dimensional continuity, momentum, and heat transport equations in the melt. The velocity field $\overrightarrow{u}$ satisfies the incompressibility condition (Eq.~\ref{eqn:1}). The momentum equation (Eq.~\ref{eqn:2}) accounts for fluid density $\rho$, pressure gradient $\nabla p$, dynamic viscosity $\mu$, gravitational acceleration $\overrightarrow{g}$, and buoyancy via thermal expansion $\alpha(T - T_0)$. The heat transport equation (Eq.~\ref{eqn:3}) incorporates specific heat $c_p$ and thermal conductivity $\lambda$ (see \cite{Rudolph2010} for a detailed formulation in crystal growth modeling).

\begin{eqnarray}
    \nabla \cdot \overrightarrow{u} = 0 \label{eqn:1}\\
    \rho \left( \frac{\partial \overrightarrow{u}}{\partial t} + (\overrightarrow{u} \cdot \nabla) \overrightarrow{u} \right) = -\nabla p + \mu \nabla^2\overrightarrow{u} - \rho \overrightarrow{g} \cdot \alpha (T - T_0) \label{eqn:2}\\
    \rho c_p \left( \frac{\partial T}{\partial t} + (\overrightarrow{u} \cdot \nabla) T \right) = \lambda \nabla^2 T \label{eqn:3}
\end{eqnarray}

Thermal boundary conditions at the solid/liquid (s/l) interface (Eqs.~\ref{eqn:4}--\ref{eqn:5}) ensure heat flux continuity and impose the melting temperature $T_m$ at the interface. Here, $\lambda_l$ and $\lambda_s$ are the thermal conductivities of the liquid and solid, $\nabla T$ the temperature gradient, $\overrightarrow{n}_{s,l}$ the interface normal, $\Delta H_s$ the latent heat of fusion, and $v$ the interface velocity.

\begin{eqnarray}
    \{ (-\lambda_l \nabla T)_l + (\lambda_s \nabla T)_s \} \cdot \overrightarrow{n}_{s,l} = \Delta H_s v \label{eqn:4} \\
    T = T_m \label{eqn:5}
\end{eqnarray}

The momentum boundary conditions assume that the velocity component normal to the solid/liquid (s/l) interface is zero, the velocity parallel to the melt-solid interface matches that of the solid, and the velocity normal to the melt-free surface is also zero.

Crystal growth was modeled using commercial software CGSim~\cite{CGSim} via step-wise axisymmetric quasi-steady-state simulations. Initial computations solved transport phenomena with a fixed interface, adjusting heater power, followed by interface shape tracking for refinement. 
Altogether, 582 datasets were generated, including 342 instances for Si, 120 instances for Ge, and 120 instances for GaAs. Some of this data has already been used in our previous studies~\cite{Dropka2024,Petkovic2025}. 

The CFD model employed a hybrid grid, combining both structured and unstructured elements. The overall grid size and the proportion of structured to unstructured elements was determined by the furnace geometry (including crystal and crucible size, as well as the amount of material being grown), which varied for each simulation case. Typically, the total grid size ranged from $10^4$ to $10^5$. Convergence was evaluated using relative residuals, with convergence considered reached when all relative residuals fell below $10^{-6}$.

\subsection{Machine Learning}\label{Sec.ML}

We employ XGBoost~\cite{Chen2016}, LightGBM~\cite{Ke2017}, and Multilayer Perceptron (MLP)~\cite{haykin1994neural} methods to model and predict interface deflection and Voronkov ratio. These models are renowned for their robust performance and efficiency in various ML tasks. They have proven effective handling small datasets and are adaptable to transfer learning scenarios~\cite{Deng2022, Zeng2019, Lopez2024}.

XGBoost and LightGBM are gradient-boosting algorithms that build models sequentially, with each iteration aiming to correct errors from the previous one. This approach enhances predictive accuracy and helps prevent overfitting, which is crucial when working with limited data. Additionally, these algorithms incorporate regularization techniques and efficient handling of missing values, further bolstering their suitability for small datasets.

Multilayer Perceptrons (MLPs) are a class of feedforward artificial neural networks that can model complex relationships within data. By implementing regularization methods such as dropout and weight decay, MLPs can effectively manage overfitting, making them appropriate for applications involving limited data.

\subsubsection{XGBoost}

XGBoost~\cite{Chen2016} (e\textbf{X}treme \textbf{G}radient \textbf{Boost}ing) is an ensemble learning method that constructs a strong predictive model by sequentially adding regression trees as weak learners. 

A \textbf{regression tree} is a type of decision tree used for predicting continuous outcomes. It works by recursively partitioning the feature space into disjoint regions and assigning a constant prediction in each region. Formally, consider a regression tree that partitions the input space into \( T \) regions \( \{ R_1, R_2, \dots, R_T \} \). The prediction for an input \( x \) is given by:
\begin{equation}
\hat{y}(x) = \sum_{j=1}^{T} c_j \, \mathbb{I}\{ x \in R_j \},
\end{equation}
where \( c_j \) is the constant prediction (often the mean of the target values) for region \( R_j \), and \( \mathbb{I}\{ \cdot \} \) is the indicator function that equals 1 if \( x \) belongs to region \( R_j \) and 0 otherwise.

The tree is built by choosing splits that minimize a loss function---typically the mean squared error (MSE) for regression tasks. For a candidate split, the algorithm evaluates the reduction in the sum of squared errors and selects the split that maximizes this reduction. The recursive nature of this process produces a tree structure that captures nonlinear relationships in the data.

XGBoost extends the idea of regression trees by employing gradient boosting, where an ensemble of trees is built sequentially. At each iteration \( t \), the model prediction for an input \( x_i \) is:
\begin{equation}
\hat{y}_i^{(t)} = \sum_{k=1}^{t} f_k(x_i), \quad f_k \in \mathcal{F},
\end{equation}
with \( \mathcal{F} \) representing the space of regression trees. The learning objective in XGBoost is defined as:
\begin{equation}
\mathcal{L}(\phi) = \sum_{i=1}^{n} \ell\big(y_i, \hat{y}_i\big) + \sum_{k=1}^{t} \Omega(f_k),
\end{equation}
where \( \ell(y_i, \hat{y}_i) \) is a differentiable loss function (e.g., squared error for regression), \( \Omega(f) \) is a regularization term that penalizes model complexity. A common form for the regularization term is:
    \begin{equation}
    \Omega(f) = \gamma T + \frac{1}{2}\alpha \|w\|^2,
   \end{equation}
where \( T \) is the number of leaves in the tree, \( w \) is the vector of leaf weights, and \( \gamma \) and \( \alpha \) are regularization parameters.

To efficiently optimize the objective, XGBoost performs a second-order Taylor expansion of the loss function at iteration \( t \):
\begin{equation}
\mathcal{L}^{(t)} \approx \sum_{i=1}^{n} \left[ g_i f_t(x_i) + \frac{1}{2} h_i f_t(x_i)^2 \right] + \Omega(f_t),
\end{equation}
with
\begin{equation}
g_i = \frac{\partial \ell\big(y_i, \hat{y}_i^{(t-1)}\big)}{\partial \hat{y}_i^{(t-1)}}, \quad h_i = \frac{\partial^2 \ell\big(y_i, \hat{y}_i^{(t-1)}\big)}{\partial \left(\hat{y}_i^{(t-1)}\right)^2}.
\end{equation}
This approximation allows XGBoost to efficiently compute the optimal structure and leaf weights for the new tree \( f_t \) using greedy algorithms.

\subsubsection{LightGBM}
LightGBM~\cite{Ke2017} (Light Gradient Boosting Machine) is a gradient-boosting framework optimized for efficiency, particularly for high-dimensional and large-scale datasets. It builds on traditional gradient boosting techniques but introduces several innovations to improve training speed and accuracy.

Unlike traditional boosting methods that evaluate all possible split points for continuous features, LightGBM uses a histogram-based approach. Instead of considering all unique values of a feature, it discretizes them into \( B \) bins, reducing computational complexity. Formally, given a feature \( x \), LightGBM maps it to a discrete bin index:
\begin{equation}
b = \text{bin}(x), \quad b \in \{1, 2, \dots, B\}.
\end{equation}
By working with bin counts rather than raw values, LightGBM significantly speeds up the training process.

Traditional boosting methods expand trees level-wise, ensuring a balanced structure. LightGBM instead employs a \textbf{leaf-wise} growth strategy, where the leaf with the highest loss reduction is split at each step. Given a candidate split \( s \) on leaf \( L \), the approximate loss reduction is given by:
\begin{equation}
\Delta \mathcal{L} \approx \frac{1}{2} \left( \frac{G_L^2}{H_L + \alpha} - \frac{G_{L_\text{L}}^2}{H_{L_\text{L}} + \alpha} - \frac{G_{L_\text{R}}^2}{H_{L_\text{R}} + \alpha} \right) - \gamma,
\end{equation}
where: \( G_L, H_L \) are the aggregated first and second-order gradients (Hessian) at leaf \( L \), \( G_{L_\text{L}}, H_{L_\text{L}} \) and \( G_{L_\text{R}}, H_{L_\text{R}} \) correspond to the left and right child nodes, while \( \alpha \) and \( \gamma \) are regularization parameters.

By selecting the optimal leaf at each step, LightGBM constructs deeper trees while maintaining computational efficiency.

\subsubsection{MLP}
A Multi-Layer Perceptron~\cite{haykin1994neural} (MLP) is a type of feed-forward neural network that learns hierarchical representations through multiple layers of non-linear transformations. It comprises an input layer, one or more hidden layers, and an output layer.

Given an input vector \( \mathbf{x} \in \mathbb{R}^{d} \), an MLP with \( L \) layers computes predictions as follows:

\begin{eqnarray}
    \mathbf{a}^{(0)} = \mathbf{x},\\
\mathbf{a}^{(l)} = \sigma\big( \mathbf{W}^{(l)} \mathbf{a}^{(l-1)} + \mathbf{b}^{(l)} \big), \quad l = 1, \dots, L,
\end{eqnarray}

where \( \mathbf{W}^{(l)} \) and \( \mathbf{b}^{(l)} \) are the weight matrix and bias vector of the \( l \)-th layer, and \( \sigma(\cdot) \) is an activation function (e.g., ReLU, sigmoid, tanh).

The final output is computed as:
\begin{equation}
\mathbf{y} = \mathbf{a}^{(L)}.
\end{equation}

The model parameters \( \{ \mathbf{W}^{(l)}, \mathbf{b}^{(l)} \} \) are trained using gradient-based optimization. The loss function \( \ell(\mathbf{y}, \mathbf{t}) \), where \( \mathbf{t} \) is the true target, is minimized using gradient descent. The gradients are computed using the backpropagation algorithm:

\begin{equation}
\frac{\partial \ell}{\partial \mathbf{W}^{(l)}} = \frac{\partial \ell}{\partial \mathbf{a}^{(L)}} \prod_{k=l+1}^{L} \frac{\partial \mathbf{a}^{(k)}}{\partial \mathbf{a}^{(k-1)}} \frac{\partial \mathbf{a}^{(l)}}{\partial \mathbf{W}^{(l)}}.
\end{equation}

\subsubsection{Hyperparameteres optimization}

To efficiently determine the optimal hyperparameters of each ML model, Bayesian Optimization (BO) was employed to explore different configurations of hidden layers and neurons while minimizing the 10-fold cross-validation mean squared error (MSE). BO, a probabilistic optimization method, models the objective function using a Gaussian Process (GP) and selects the next evaluation points via an acquisition function, balancing exploration and exploitation \cite{Shahriari2016, Snoek2012}. 

For MLP, the search space included architectures with 1 to 3 hidden layers, with neuron counts constrained to a maximum of 32 per layer, ensuring manageable computational complexity. The optimization considered two structural patterns: funnel architectures, where neurons decrease toward a latent bottleneck before increasing, and monotonic architectures, where neuron counts either consistently increase or decrease from input to output. Activation functions were selected dynamically for each layer from ReLU, Tanh, Sigmoid, and LeakyReLU, while dropout rates varied between 0.01 and 0.5 to regulate overfitting. Additionally, optimizers were chosen from Adam, RMSprop, and SGD, with learning rates ranging from 1e-4 to 1e-2.

In the case of LightGMB, the search space included key hyperparameters such as learning rate, which was optimized on a log-uniform scale between 0.001 and 0.1; the number of leaves controlling tree complexity (ranging from 20 to 200); and the number of estimators, tuned between 50 and 300. Additionally, the subsample ratio, which regulates feature and data sampling during training, was varied between 0.5 and 1.0 to enhance generalization.

The optimization also employed 10-fold cross-validation to ensure robust evaluation, minimizing the negative mean squared error (MSE) as the objective function. A total of 50 BO iterations were conducted, balancing computational efficiency with search precision.

The optimization process included early stopping and monitoring validation loss during training to prevent overfitting and improve generalization to unseen data \cite{Prechelt1998}. 

\subsection{Transfer Learning}
\label{sec:transferlearning}
Transfer learning~\cite{Pan2010,Weiss2016} is a machine learning paradigm that enables knowledge transfer from a source domain \( D_s \) to a target domain \( D_t \), where the two domains may differ in their feature spaces or data distributions. This approach is particularly useful when the target domain has limited data, as it allows models to leverage knowledge from a related source domain with abundant data. In crystal growth, transfer learning can help bridge the gap between materials with different properties, such as thermal conductivity, viscosity, and surface tension.

The core idea of transfer learning is to address the discrepancy between the source and target domains, which can be formalized as:
\begin{equation}
P(X_s) \neq P(X_t) \quad \text{or} \quad P(Y_s | X_s) \neq P(Y_t | X_t),
\end{equation}
where \( X_s \) and \( X_t \) represent the feature spaces of the source and target domains, respectively, and \( Y_s \) and \( Y_t \) are their corresponding outputs. The goal is to learn a model \( f \) that performs well on the target domain by leveraging knowledge from the source domain.

A general formulation of the transfer learning optimization problem is:

\begin{equation}
\theta^* = \arg\min_{\theta} \sum_{i \in S} L_s(f(X_i; \theta), Y_i) + \alpha \sum_{j \in T} L_t(f(X_j; \theta), Y_j),
\end{equation}
where \( \theta \) represents the model parameters, \( L_s \) and \( L_t \) are the loss functions for the source and target domains, respectively, and \( \alpha \) is a trade-off parameter that balances the contributions of the source and target domains \cite{Pan2010, Weiss2016}.
This formulation ensures that the model learns from both domains while prioritizing performance on the target domain.

\subsubsection{Warm Start Transfer} Warm Start (WS) is a technique in machine learning that utilizes a pre-trained model by initializing its weights from a previously trained network, followed by fine-tuning on new data. WS allows the model to adapt to the target domain while retaining learned representations from the source domain. The initial training on a larger or related dataset enables the model to capture a diverse range of features and patterns, which are crucial for effective transfer learning. With its informed weight initialization, the pre-trained network is a robust starting point for further training, allowing the model to optimize from a more advantageous position in the parameter space~\cite{Kornblith2019, Gao2016, Guo2019}.

The advantages of WS are particularly evident in its ability to accelerate convergence. By starting with well-informed weights, the model can navigate the parameter space more efficiently and significantly reduce the number of training epochs. Research has shown that models initialized with pre-trained weights converge faster than those initialized randomly, as they begin the optimization process closer to an optimal solution~\cite{Gao2016, Guo2019}. Moreover, the pre-trained weights encapsulate meaningful patterns and domain-specific knowledge, enhancing the stability of the training process. This stability is crucial, especially when dealing with small or noisy target datasets, as it mitigates the risk of the model becoming trapped in suboptimal local minima. In fact, by starting near a favorable region in the loss landscape, WS helps steer the optimization away from the multitude of local minima, further enhancing convergence~\cite{Guo2020, Hong2021}.

In addition to enhancing convergence and stability, the warm start technique serves as an implicit regularizer. The model is less prone to overfitting since the the optimization process is centered around previously learned representations. This is especially advantageous when the source and target domains exhibit similarities, as the pre-trained model can transfer valuable insights from the source domain to the target domain. In such scenarios, fine-tuning primarily files existing representations, ensuring that the adaptation to the target domain is both efficient and effective~\cite{Dong2021, Tajbakhsh2016}. 

Overall, Warm Start Transfer represents a practical strategy for knowledge transfer between related domains. By leveraging the benefits of accelerated convergence, improved training stability, and an inherent regularization effect, WS facilitates the development of models that are well-adapted to new tasks while capitalizing on the robust features learned from prior training experiences~\cite{Kornblith2019, Guo2020, Hong2021}. This methodology enhances the efficiency of model training and contributes to the robustness and generalizability of machine learning models across diverse applications.

\subsubsection{ Merged Training}
Merged training (MT) involves jointly training a model on both the source and target domains, which allows the model to generalize across both domains. The loss function represents a weighted sum of the source and target losses:
\begin{equation}
L(\theta) = \alpha L_s(\theta) + (1 - \alpha) L_t(\theta),
\end{equation}
where \( \alpha \) controls the relative importance of the source and target domains. This approach is advantageous when the target domain has sufficient data to contribute meaningfully to the training process \cite{goodfellow2016}. By training on a broader dataset, the model gains exposure to diverse patterns, making it more robust to variations in experimental conditions while reducing the risk of overfitting to a specific domain. 

\subsubsection{Hyperparameters Transfer}
Hyperparameters Transfer (HPT) strategy involves selecting optimal hyperparameters from a previously trained model while training a new model from scratch on the target dataset. Unlike Warm Start Transfer, no weight transfer occurs. HT is practical when the source and target domains are too dissimilar for direct parameter transfer. The optimization problem can be written as:
\begin{equation}
\theta^* = \arg\min_{\theta} L_t(f(X_t; \theta), Y_t),
\end{equation}
where \( \theta \) is initialized using hyperparameters tuned on the source domain. 
The main advantages are efficient model optimization by reusing previously tuned hyperparameters, reducing computational costs, adaptability, and allowing independent training on new datasets while maintaining a well-tuned architecture. This approach is also useful when direct knowledge transfer via pre-trained weights is not feasible, but optimized hyperparameter selection remains beneficial.

\subsection{Data analysis and models evaluation}\label{sec:dataanalysis}
In this study, we simulated the growth of three different materials: Silicon (Si), Germanium (Ge), and Gallium Arsenide (GaAs). 
We created 582 datasets in total, including 342 datsets for Si, and 120 data sets Ge and GaAs, respectively.
To quantify data sets similarities we use feature-wise descriptive statistics, namely mean and variance given by:

\[ \mu = \frac{\sum_{n=1}^{N} x_n }{N}\]

\[ \sigma^2 = \frac{\sum_{n=1}^{N} \mu-x_n }{N},\]

where $N$ is a number of instances. 
Additionally, we calculate Maximum Mean Discrepancy (MMD) - a statistical measure used to quantify the difference between two probability distributions \(P\) and \(Q\) based on samples drawn from each. Given two datasets \(\{x_i\}_{i=1}^m\) from \(P\) and \(\{y_j\}_{j=1}^n\) from \(Q\), and a suitable kernel function \(k(\cdot,\cdot)\) (often a Gaussian RBF or polynomial kernel~\cite{Hofmann2008}), the empirical MMD is defined as:\\

\begin{equation}
    \mathrm{MMD}^2(P, Q) 
= \frac{1}{m(m-1)} \sum_{i \neq i'} k\bigl(x_i, x_{i'}\bigr) 
+ \frac{1}{n(n-1)} \sum_{j \neq j'} k\bigl(y_j, y_{j'}\bigr)
- \frac{2}{mn} \sum_{i, j} k\bigl(x_i, y_j\bigr).
\end{equation}

Intuitively, \(\mathrm{MMD}(P, Q)\) will be small if the two distributions \(P\) and \(Q\) overlap significantly (are similar) and will be larger if they are dissimilar. Because MMD leverages kernel functions, it can capture high-dimensional and nonlinear structures in the data, making it particularly useful for comparing complex distributions.

The ML models accuracy was evaluated using two key error metrics, root mean square error (RMSE) and mean absolute error (MAE):

\[
\mathrm{RMSE} = \sqrt{\frac{1}{n} \sum_{i=1}^{n} (y_i - \hat{y}_i)^2},
\]
\[
\mathrm{MAE} = \frac{1}{n} \sum_{i=1}^{n} \left|y_i - \hat{y}_i\right|,
\]

where \(y_i\) represents the true values, \(\hat{y}_i\) represents the predicted values, and \(n\) is the number of instances in the dataset. %While RMSE measures the square root of the average of the squared differences between actual and predicted values, giving more weight to larger errors, MAE measures the average absolute difference between predicted and true values. It treats all errors equally, providing a more interpretable measure of the average error magnitude.

Additionally, we calculate skill scores to compare different methods and quantify the improvement of models' ability to transfer learned knowledge across materials.
Skill scores are defined as the relative improvement of a model compared to a benchmark model and are calculated as:
\[
\text{Skill Score} = 1 - \frac{\mathrm{RMSE}_{\text{model}}}{\mathrm{RMSE}_{\text{baseline}}}.
\]
A higher skill score indicates better performance relative to the benchmark, while a negative score indicates performance worse than the baseline. In the case of equal performance, the skill score is equal to zero, while perfect prediction will give a value of 1.

\section{Results and Discussion}
\label{sec:results}
\begin{figure}
\centering
  \includegraphics[width=0.5\linewidth]{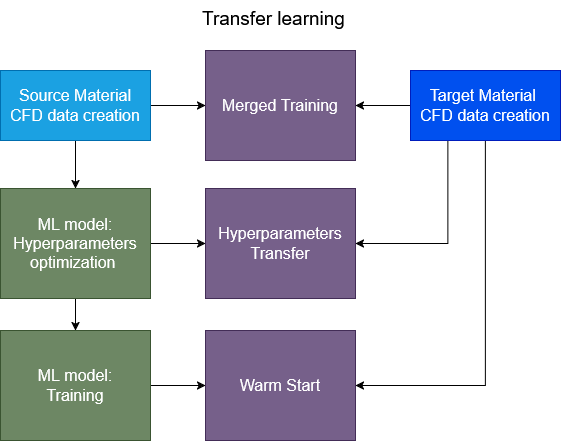}
  \caption{The architecture of transfer learning in semiconductors Cz-crystal growth}
  \label{fig:architecture}
\end{figure}

This study investigates the application of three transfer learning techniques, Warm Start Transfer (WS), Merged Training (MT), and Hyperparameter Transfer(HPT), to predict interface deflection $\Delta z$ and Voronkov criterion $v/G$ in semiconductor materials, specifically Silicon (Si), Germanium (Ge), and Gallium Arsenide (GaAs). We employ three machine learning models: LightGBM, XGBoost, and MLP. 
The architecture of the proposed approach is shown in Figure \ref{fig:architecture}.

The statistics of data sets for different growth materials are given in Table~\ref{tab:summary_statistics}

\begin{table}[ht!]
    \centering
    \small
    \begin{tabular}{c|lll|lll}
    \hline
    \hline 
         & \multicolumn{3}{c|}{Mean} & \multicolumn{3}{c}{Variance} \\
         \hline 
        \textbf{Feature} & \textbf{Si} & \textbf{Ge} & \textbf{GaAs} & \textbf{Si} & \textbf{Ge} & \textbf{GaAs} \\
        \hline \hline
        $x_1$  & 15.73  & 7.58   & 7.63   & 382.52 & 6.12    & 4.62    \\
        $x_2$  & 24.53  & 25.32  & 24.86  & 185.49 & 181.46  & 192.72  \\
        $x_3$  & 9.77   & 6.15   & 6.21   & 15.24  & 1.28    & 1.46    \\
        $x_4$  & 5.12   & 2.95   & 3.19   & 5.80   & 0.94    & 1.03    \\
        $x_5$  & 28.17  & 32.28  & 31.25  & 244.30 & 517.84  & 504.75  \\
        $x_6$  & 117.68 & 156.16 & 161.54 & 207.47 & 1065.41 & 2323.51 \\
        $x_7$  & 155.01 & 158.11 & 153.22 & 435.71 & 1032.18 & 892.65  \\
        $x_8$  & 19.86  & 26.36  & 29.12  & 36.29  & 153.61  & 170.61  \\
        $x_9$  & 0.70   & 0.57   & 0.67   & 0.04   & 0.06    & 0.04    \\
        $x_{10}$ & 25.30  & 18.40  & 24.45  & 220.38 & 591.34  & 674.85  \\
        $x_{11}$ & 0.66   & 0.38   & 7.99   & 0.19   & 0.06    & 16.97   \\
        $x_{12}$ & 12.77  & 28.04  & 11.90  & 57.78  & 291.86  & 38.91   \\
        $x_{13}$ & -3.77  & -6.85  & -7.51  & 6.67   & 37.91   & 30.49   \\
        $x_{14}$ & 25.66  & 2.41   & 12.48  & 207.45 & 0.71    & 71.82   \\
        $x_{15}$ & 32.62  & 1.86   & 19.29  & 301.86 & 0.90    & 73.21   \\
        \hline \hline
    \end{tabular}
    \caption{Descriptive statistics for Si, Ge, and GaAs datasets.}
    \label{tab:summary_statistics}
\end{table}

The descriptive statistics reveal both broad similarities and notable discrepancies among the three datasets. Several features, such as $x_2$ and $x_3$, maintain relatively stable mean values and variances, indicating consistent behaviour across conditions. For instance, $x_3$ exhibits minor variances (15.24, 1.28, and 1.46) despite a moderately elevated mean in the Si dataset.

By contrast, features like $x_1$ and $x_{14}$ display considerable fluctuations. In particular, $x_{14}$ shows a marked change in mean values (25.66, 2.41, 12.48) coupled with drastically different variances (207.45, 0.71, 71.82), signaling substantial shifts in both location and spread across the datasets.

\begin{table}[ht!]
\centering

\begin{tabular}{c|cc}
\hline \hline
\textbf{Variable}  & \textbf{MMD (Si vs Ge)} & \textbf{MMD (Si vs GaAs)} \\ \hline
$x_1$              & 0.5497                              & 0.7065                                    \\
$x_2$              & 0.0997                              & 0.1090                                    \\
$x_3$              & 0.2522                              & 0.2494                                    \\
$x_4$              & 0.1998                              & 0.1763                                    \\
$x_5$              & 0.2336                              & 0.2205                                    \\
$x_6$              & 0.3881                              & 0.4710                                    \\
$x_7$              & 0.1742                              & 0.1525                                    \\
$x_8$              & 0.2467                              & 0.2540                                    \\
$x_9$              & 0.0167                              & 0.0009                                    \\
$x_{10}$           & 0.4045                              & 0.4377                                    \\ \hline
$x_{11}$           & 0.0513                              & 0.9923                                    \\
$x_{12}$           & 0.1072                              & 0.0443                                    \\
$x_{13}$           & 0.0280                              & 0.0902                                    \\
$x_{14}$           & 0.7290                              & 0.3419                                    \\
$x_{15}$           & 0.6583                              & 0.0590                                    \\ \hline
\textbf{Total MMD} & \textbf{0.0115}                     & \textbf{0.0116}    \\ \hline \hline                   
\end{tabular}

 \caption{Feature-wise and total MMD values for two target data sets (Ge and GaAs) compared to source dataset (Si).}  
\label{tab:mmd}
\end{table}

We calculate MMD~\ref{sec:dataanalysis} for each individual feature to assess how different the target material distributions (Germanium or Gallium Arsenide) are from the source material (Silicon). Table~\ref{tab:mmd} reports the MMD value for each feature, along with a total MMD that aggregates the distributional difference across all features. Notably, certain features (e.g., $x_{14}$  for Si vs Ge, $x_{11}$ for Si vs GaAs) exhibit high MMD values, indicating substantial differences in those aspects of the material. Other features like $x_9$ and $x_{15}$ have low MMD values for GaAs, suggesting closer alignment with Si in those dimensions.

The total MMD of approximately \(0.0115\) for both Ge and GaAs indicates that, at a global level, the differences between Silicon and the target materials are not overly significant. However, the per-feature breakdown reveals more nuanced disparities. In the context of transfer learning, these MMD values help identify which features may need domain adaptation or more careful model calibration. If certain features are heavily mismatched, a simple parameter transfer is more prone to fail, motivating methods such as weighted feature alignment or adversarial domain adaptation. In our experiments, aligning or otherwise accounting for features with notably high MMD values proved crucial for improving predictive performance when transferring knowledge from Silicon to either Germanium or Gallium Arsenide.

\begin{table}
\centering
\begin{tabular}{l|llll|llll}
\hline \hline
\multirow{2}{*}{} & \multicolumn{4}{c|}{$\Delta z$}                              & \multicolumn{4}{c}{$v/G$}                                         \\  \hline
                  & \multicolumn{2}{c}{RMSE} & \multicolumn{2}{c|}{MAE} & \multicolumn{2}{c}{RMSE ($\times10^3$)} & \multicolumn{2}{c}{MAE($\times10^3$)} \\  
Model             & Train       & Test       & Train      & Test       & Train          & Test          & Train         & Test          \\  \hline
LightGMB          & 2.755       & 6.891      & 1.682      & 4.262      & 0.250          & 0.465         & 0.131         & 0.260         \\
XGBoost           & 1.739       & 6.823      & 1.048      & 4.168      & 0.311          & 0.459         & 0.200         & 0.275         \\
MLP               & 3.710       & 6.817      & 2.350      & 4.255      & 0.551          & 0.584         & 0.409         & 0.443       \\
 \hline  \hline
\end{tabular}

\caption{$\Delta z$ and $v/G$ error metrics (RMSE and MAE) for Silicon (Si) across models.}
\label{tab:silicon}
\end{table}

The results of the Silicon base case, accuracy of the ML models (average RMSE and MSE on 10-fold cross-validation) predicting Silicon growth output parameters ($\Delta z$ and $v/G$) are shown in Table~\ref{tab:silicon}. 
While in predicting $\Delta z$ all three models show similar accuracy on the test set, in the case of $v/G$ XGBoost and LightGBM outperform the MLP model significantly. Despite extensive optimization of the MLP architecture using Bayesian optimization, its performance remains suboptimal, suggesting that factors other than the model architecture, such as insufficient training data, may influence its performance. 
%In the Silicon baseline for $v/G$, LightGBM and XGBoost exhibit similarly strong performance, with XGBoost achieving slightly lower RMSE but near-equivalent MAE values. MLP, despite being extensively optimized, displays higher RMSE and MAE. Unlike in the deflection task, the gap between MLP and the tree-based models remains noticeable. Still, it is not as large, suggesting that the ratio $v/G$ may be somewhat simpler to model with a feed-forward neural network than the deflection variable. Nevertheless, MLP still lags in capturing the underlying patterns in Silicon compared to XGBoost and LightGBM.

Next, we trained and tested the ML models on Ge and GaAs data sets and compared the performance to three transfer learning approaches (see Section~\ref{sec:transferlearning}) where the model created on the Si data set was used as a source model. The training/ test results are again obtained by 10-fold cross-validation, and the ML models' hyperparameters were optimized with Bayesian optimization.

\begin{table}
\centering
\scalebox{0.9}{
\begin{tabular}{l|llll|llll}
\hline  \hline
\multirow{2}{*}{}        & \multicolumn{4}{c|}{$\Delta z$}                                   & \multicolumn{4}{c}{$v/G$}                                         \\  \hline
                         & \multicolumn{2}{c}{RMSE}        & \multicolumn{2}{c|}{MAE}        & \multicolumn{2}{c}{RMSE($\times10^3$)}        & \multicolumn{2}{l}{MAE($\times10^3$)}         \\
Model                    & Train          & Test           & Train          & Test           & Train          & Test           & Train          & Test           \\ \hline
LightGMB                 & 1.494          & 2.778          & 0.974          & 2.161          & \textbf{0.184} & \textbf{0.339} & \textbf{0.111} & \textbf{0.262} \\
Warm Start               & 0.574          & 3.671          & 0.422          & 2.696          & 0.039          & 0.292          & 0.018          & 0.213          \\
Merged Training          & 0.812          & 3.164          & 0.631          & 2.270          & \textbf{0.026} & 0.279          & 0.018          & \textbf{0.202} \\
Hyperparameters Transfer & 0.393          & 2.474          & 0.300          & 1.929          & 0.037          & \textbf{0.276} & \textbf{0.017} & 0.218          \\ \hline
XGBoost                  & \textbf{0.731} & \textbf{2.563} & \textbf{0.512} & \textbf{2.007} & 0.276          & 0.339          & 0.209          & 0.267          \\
Warm Start               & 0.017          & 2.390          & 0.011          & 1.864          & 0.261          & 0.343          & 0.201          & 0.271          \\
Merged Training          & 0.551          & 3.221          & 0.393          & 2.231          & 0.249          & 0.325          & 0.190          & 0.250          \\
Hyperparameters Transfer & \textbf{0.016} & \textbf{2.374} & \textbf{0.010} & \textbf{1.836} & 0.261          & 0.341          & 0.201          & 0.271          \\ \hline
MLP                      & 2.783          & 4.110          & 2.035          & 2.975          & 0.989          & 0.967          & 0.754          & 0.803          \\
Warm Start               & 2.308          & 2.494          & 1.724          & 1.936          & 0.459          & 0.612          & 0.364          & 0.519          \\
Merged Training          & 0.779          & 2.805          & 0.579          & 2.187          & 1.697          & 1.538          & 1.382          & 1.182          \\
Hyperparameters Transfer & 2.779          & 3.171          & 2.051          & 2.710          & 10.897         & 11.528         & 8.873          & 9.8063         \\ \hline  \hline
\end{tabular}
}
\caption{Average error metrics (RMSE and MAE) for $\Delta z$ and $v/G$ for Germanium (Ge) across transfer learning approaches. }
\label{tab:ge_deflection_vg}
\end{table}

As detailed in Table~\ref{tab:ge_deflection_vg}, in a case of predicting interface deflection $\Delta z$, a noticeable improvement is observed across all models when transferring from Silicon to Germanium. The XGBoost model has the lowest errors when using only the Ge data set and also achieves the best results on the test set with transfer learning, with an average RMSE of 2.374 and an average MAE of 1.836. The benefit of transfer learning is also noticeable with the MLP model, even though the accuracy remains slightly lower than in the tree-based models.

When transferring from Silicon to Germanium (see Table~\ref{tab:ge_deflection_vg}) in predicting $v/G$, we observe that both XGBoost and LightGBM enjoy notable reductions in error through Hyperparameters Transfer, similar to the $\Delta z$ prediction results. LightGBM (MT), for instance, substantially lowers its RMSE and MAE relative to the baseline (LightGBM with no knowledge transfer), demonstrating that incorporating Silicon data alongside Germanium improves generalization. By contrast, MLP’s baseline performance in Germanium is already relatively poor (with comparatively high RMSE and MAE). While Warm Start Transfer reduces these errors, they remain higher than those of either tree-based model. A particularly striking case is MLP with the Hyperparameter Transfer approach, where parameter mismatches can lead to significant errors, indicating that naive parameter reuse is less effective for MLP when the data distributions differ.

\begin{table}[]
\centering
\scalebox{0.9}{
\begin{tabular}{l|llll|llll}
\hline \hline
\multirow{2}{*}{}        & \multicolumn{4}{c|}{$\Delta z$}                                            & \multicolumn{4}{c}{$v/G$}                                            \\ \hline
                         & \multicolumn{2}{c}{RMSE}        & \multicolumn{2}{c|}{MAE}        & \multicolumn{2}{c}{RMSE($\times10^3$)}        & \multicolumn{2}{l}{MAE($\times10^3$)}         \\
Model                    & Train          & Test           & Train          & Test           & Train          & Test           & Train          & Test           \\ \hline
LightGMB                 & \textbf{2.893} & 5.621          & \textbf{1.817} & \textbf{4.062} & 2.600          & 4.066          & 1.515          & \textbf{2.732} \\
Warm Start               & 1.149          & 5.254          & 0.862          & 3.889          & 1.478          & 3.315          & 0.802          & 2.248          \\
Merged Training          & 1.287          & 5.154          & 0.957          & 3.809          & \textbf{1.418} & 3.052          & 0.781          & 2.037          \\
Hyperparameters Transfer & 1.124          & 4.776          & 0.797          & 3.328          & 1.502          & 3.360          & 0.795          & 2.231          \\ \hline
XGBoost                  & 4.225          & \textbf{4.894} & 3.432          & 4.145          & \textbf{1.514} & \textbf{3.497} & \textbf{0.795} & 2.290          \\
Warm Start               & 1.039          & 4.687          & 0.633          & 3.162          & 0.860          & 3.428          & 0.538          & 2.249          \\
Merged Training          & 1.498          & 5.138          & 1.077          & 3.794          & \textbf{0.833} & \textbf{2.869} & 0.580          & \textbf{2.028} \\
Hyperparameters Transfer & \textbf{1.021} & 4.647          & \textbf{0.628} & 3.183          & 0.864          & 3.521          & \textbf{0.530} & 2.343          \\ \hline
MLP                      & 4.317          & 5.293          & 3.432          & 4.427          & 5.783          & 7.640          & 2.852          & 4.002          \\
Warm Start               & 3.175          & \textbf{3.640} & 2.224          & \textbf{3.105} & 4.244          & 3.203          & 3.198          & 2.667          \\
Merged Training          & 9.656          & 8.074          & 7.022          & 6.467          & 7.612          & 5.420          & 5.670          & 4.408          \\
Hyperparameters Transfer & 2.495          & 4.881          & 1.430          & 3.322          & 6.307          & 4.320          & 4.802          & 3.658          \\\hline \hline   
\end{tabular}
}
\caption{Average error metrics (RMSE and MAE) for $\Delta z$ and $v/G$  for Gallium Arsenide (GaAs)  across transfer learning approaches. }
\label{tab:gaas_deflection_vg}
\end{table}

The Gallium Arsenide results underscore similar trends (see Table~\ref{tab:gaas_deflection_vg}). For $\Delta z$ XGBoost’s baseline errors are relatively high but drop markedly with Merged Training.

MLP again demonstrates the greatest variability in outcomes, performing significantly better with Warm Start Transfer while achieving the lowest errors overall (RMSE of 3.64, MAE of 3.105). These findings underscore MLP’s higher sensitivity to data quantity and initialization strategy, emphasizing that, while neural networks can benefit from transfer learning, they may require more carefully curated approaches and larger datasets to match the resilience of tree-based models.
On the other hand, in the case of $v/G$, XGBoost model shows the best performance both with and without transfer learning.

\begin{figure}[ht!]
    \centering
    \begin{subfigure}[b]{0.8\textwidth}
        \centering
        \includegraphics[width=\textwidth]{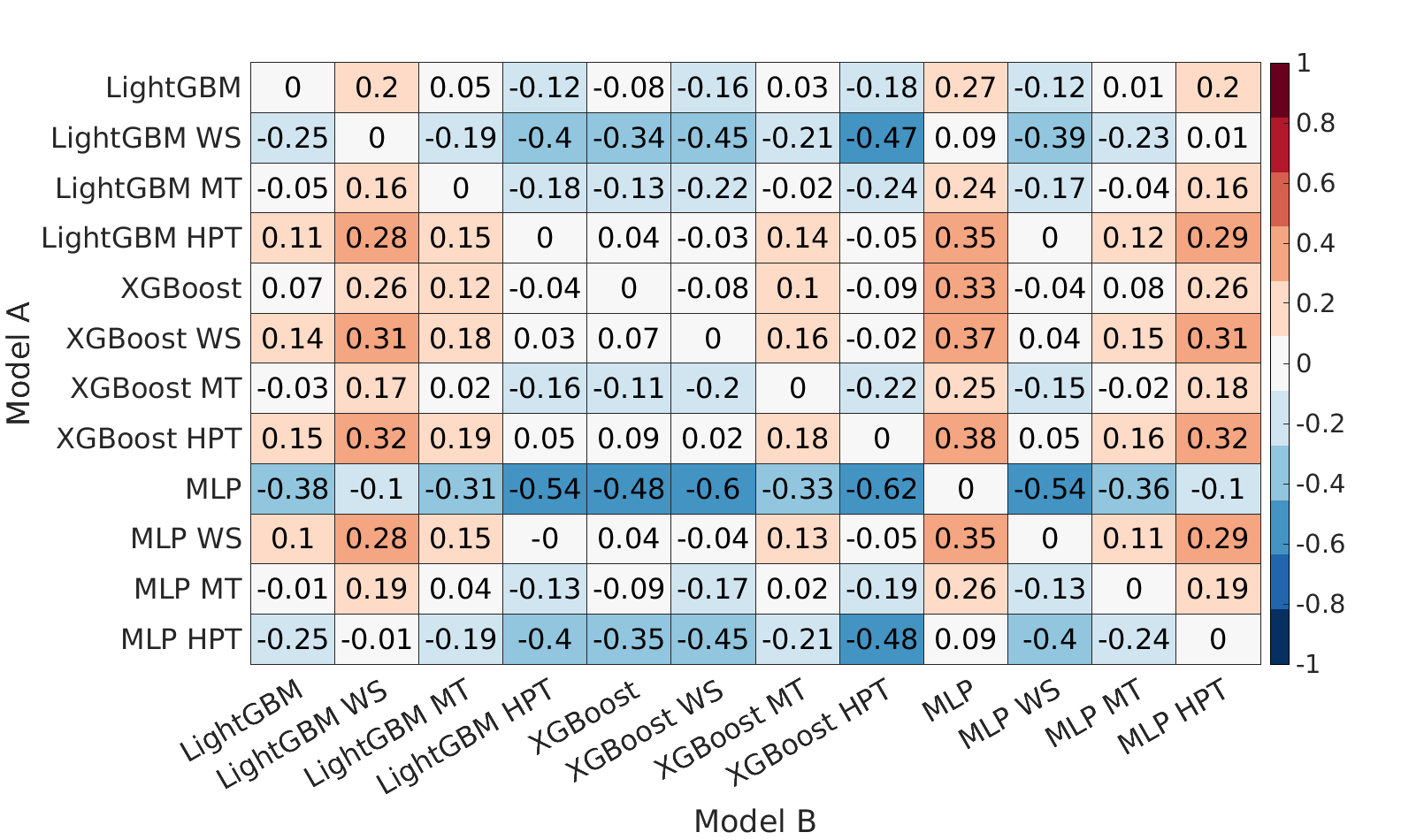}
        \caption{ $\Delta z$ model comparison}
        \label{fig:skillscore_deflection_ge}
    \end{subfigure}
    \hfill
    \begin{subfigure}[b]{0.78\textwidth}
        \centering
        \includegraphics[width=\textwidth]{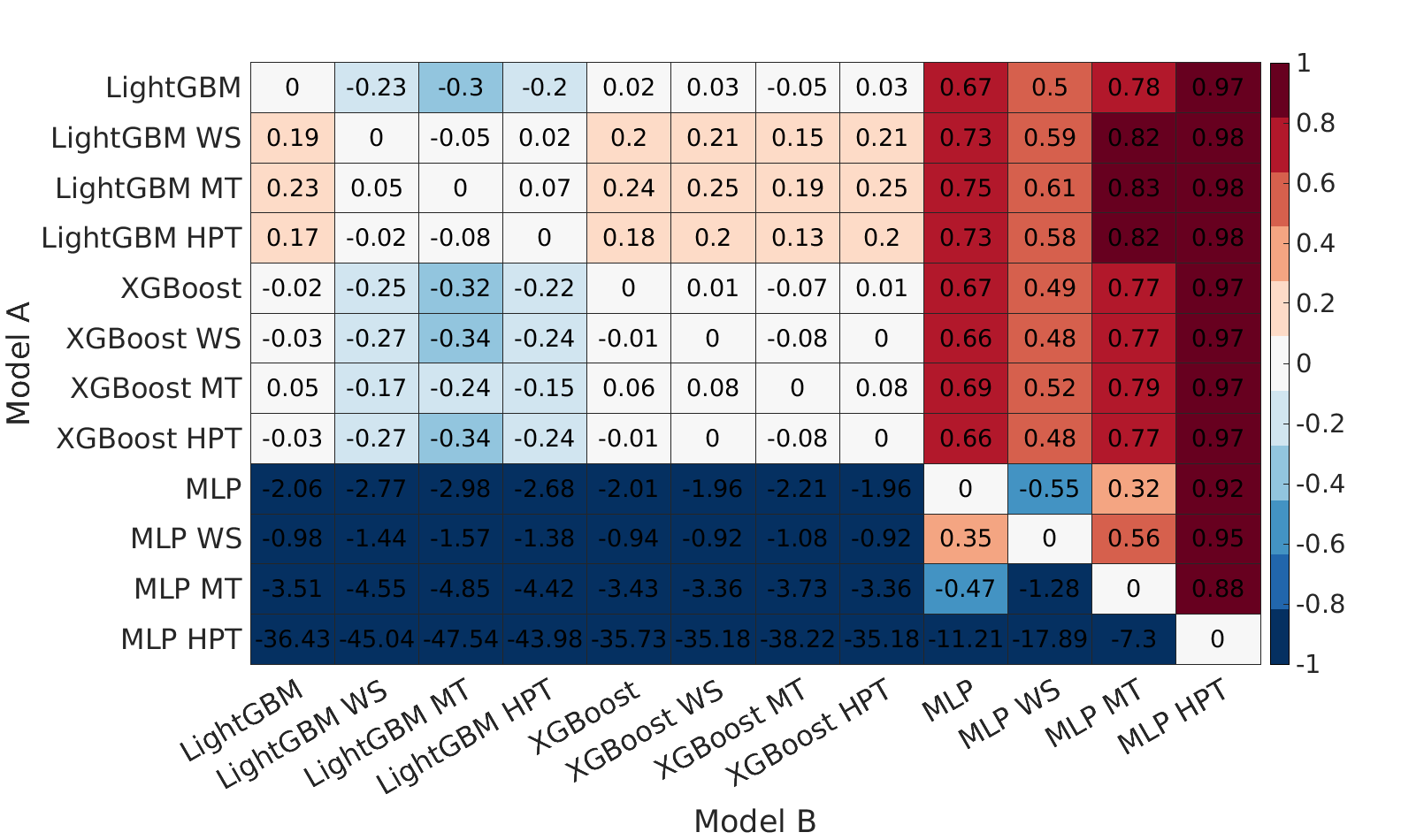}
        \caption{ $v/G$ models comparison}
        \label{fig:skillscore_vg_ge}
    \end{subfigure}
     
    \caption{ Skill score heatmap for the different models and transfer learning techniques for predicting $\Delta z$ and $v/g$ on Germanium. }
    \label{fig:skillscore_ge}
\end{figure}

\begin{figure}[ht!]
    \centering
    \begin{subfigure}[b]{0.8\textwidth}
        \centering
        \includegraphics[width=\textwidth]{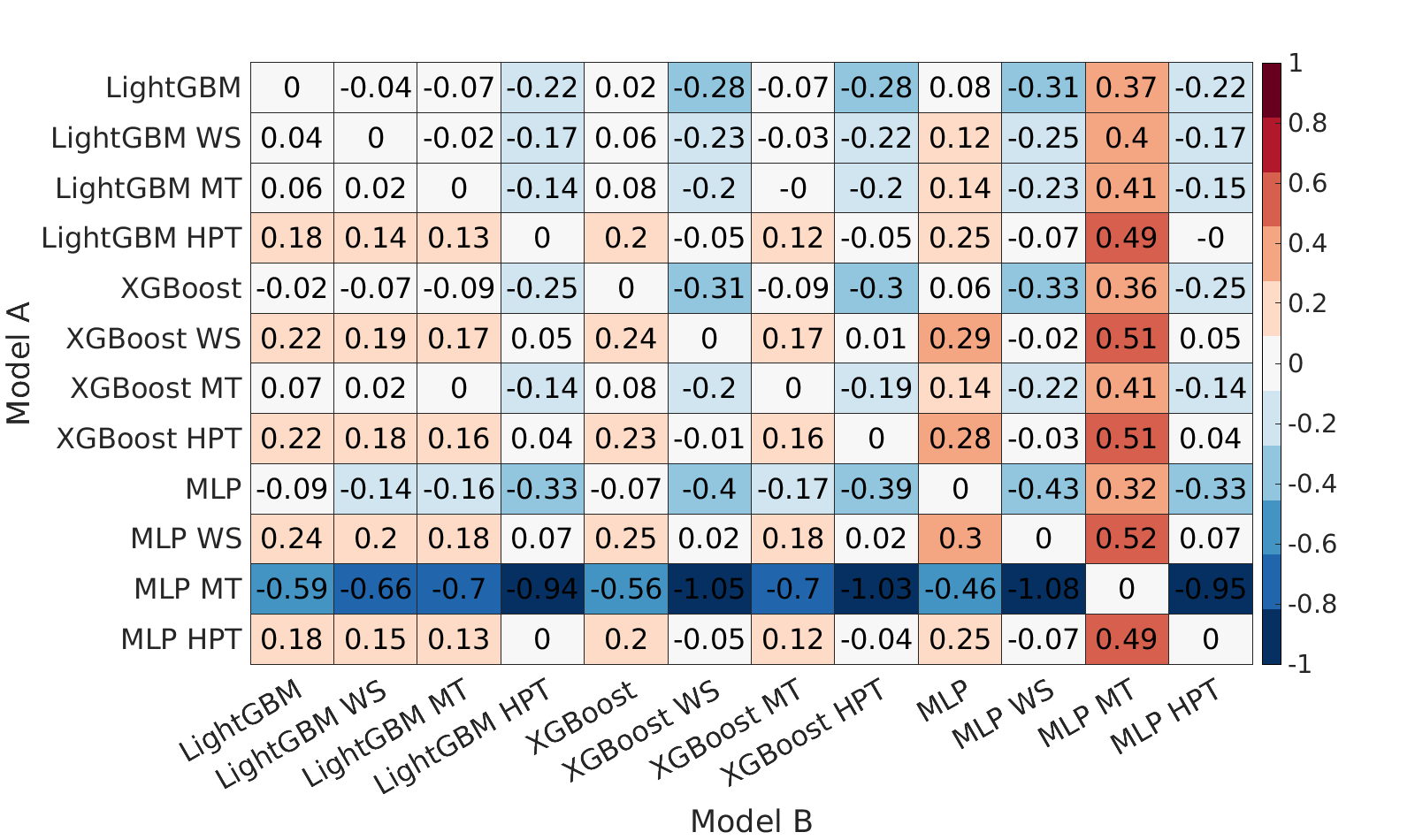}
        \caption{$\Delta z$ model comparison}
        \label{fig:skillscore_deflection_gaas}
    \end{subfigure}
    \hfill
    \begin{subfigure}[b]{0.78\textwidth}
        \centering
        \includegraphics[width=\textwidth]{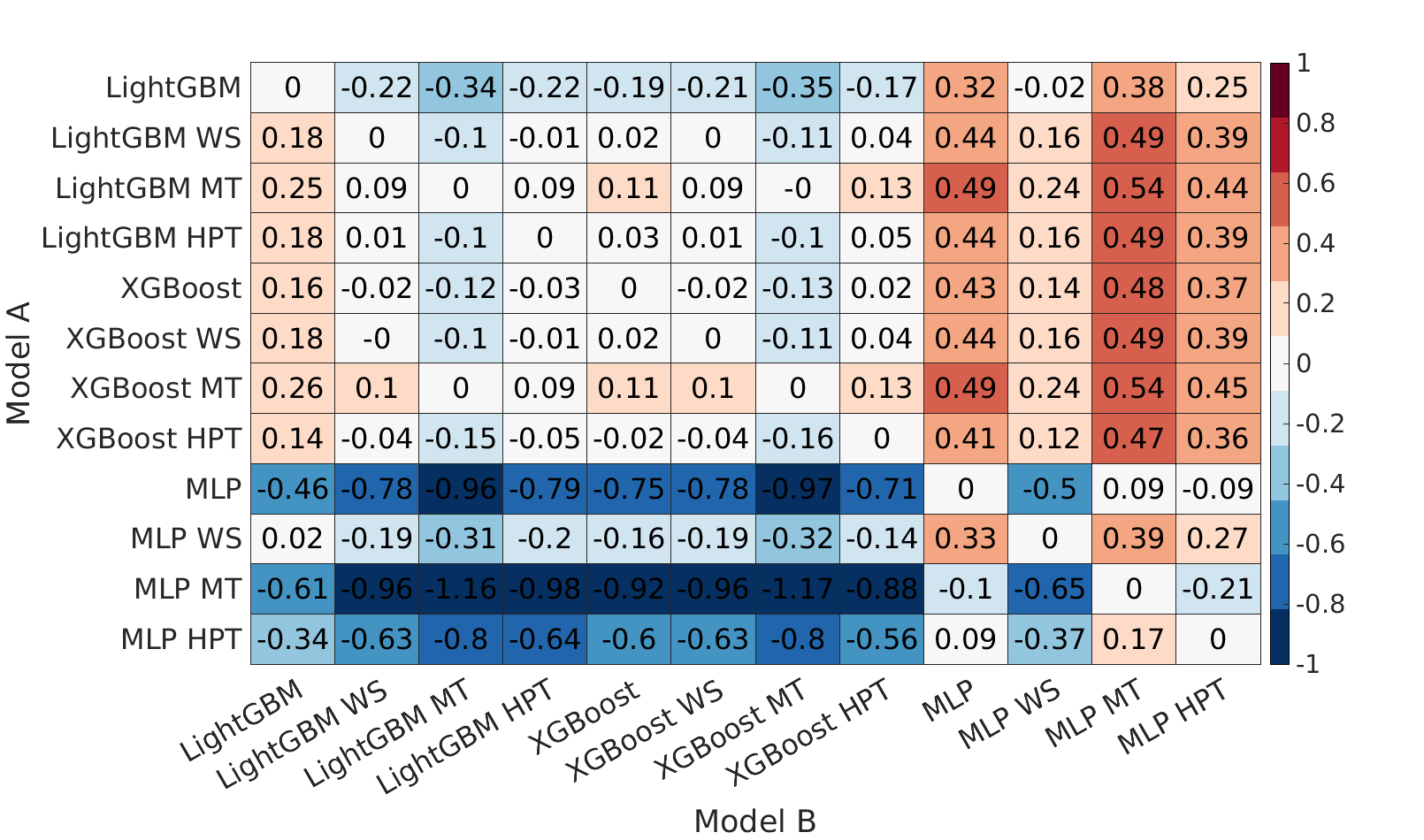}
        \caption{ $v/G$ models comparison}
        \label{fig:skillscore_vg_gaas}
    \end{subfigure}
     
     \caption{ Skill score heatmap for the different models and transfer learning techniques for predicting $\Delta z$ and $v/g$ on Gallium Arsenide. }
    \label{fig:skillscore_gaas}
\end{figure}

The skill scores, illustrated in Figures~\ref{fig:skillscore_ge} and~\ref{fig:skillscore_gaas}, further emphasize our findings. For Germanium and predicting $\Delta z$ (Figure~\ref{fig:skillscore_deflection_ge}) XGBoost delivers the highest skill scores across all transfer learning methods when using Hyperparameters Transfer (HPT), outperforming other models for up to 38\%. LightGMB also performs well with HPT, achieving skill scores around 0.35 compared to MLP models. However, MLP shows negative skill scores compared to the other two methods but still benefits from transfer learning by improving its performance by 35\% with the Warm Start transfer learning approach. Figure~\ref{fig:skillscore_vg_ge} again shows dominant performance of LightGBM model, especially with WS and MT transfer learning. MLP continues to underperform, indicating that even with transfer learning techniques, MLP struggles to adapt to Germanium.

Skill scores for different models for predicting output growth parameters for Gallium Arsenide are shown in Figure~\ref{fig:skillscore_gaas}). In this case, when predicting $\Delta z$ (Figure~\ref{fig:skillscore_deflection_gaas}), LightGMB and XGBoost models show similar performance with improving the accuracy for around 20\% when using HPT. While in the base case, MLP is still outperformed by tree-based models, transfer learning MLP (WS) achieves the overall best result. 

Predicting $v/G$ for GaAs shows different results (Figure~\ref{fig:skillscore_vg_gaas}).
Both tree-based models overperform MLP and all MLP transfer learning models for 12\% to 49\%. With the skill score of 0.49 compared to base case MLP Hyperparameters Transfer is the best model for predicting $v/G$ in GaAs crystal growth.

\begin{figure}[ht!]
    \centering
    \begin{subfigure}[b]{0.48\textwidth}
        \centering
        \includegraphics[width=\textwidth]{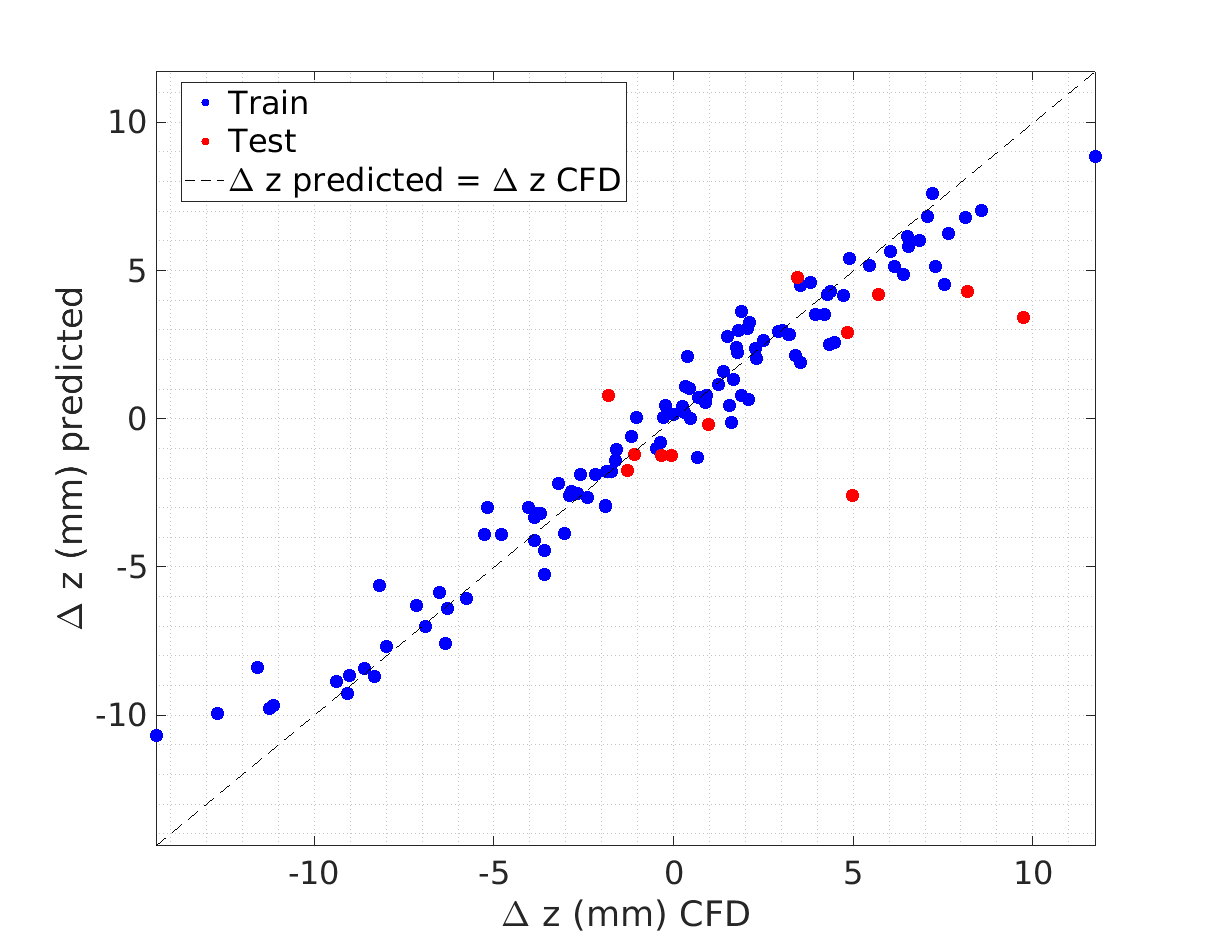}
        \caption{XGBoost model for Ge $\Delta z$ prediction}
        \label{fig:XGB_def_Ge}
    \end{subfigure}
    \hfill
    \begin{subfigure}[b]{0.48\textwidth}
        \centering
        \includegraphics[width=\textwidth]{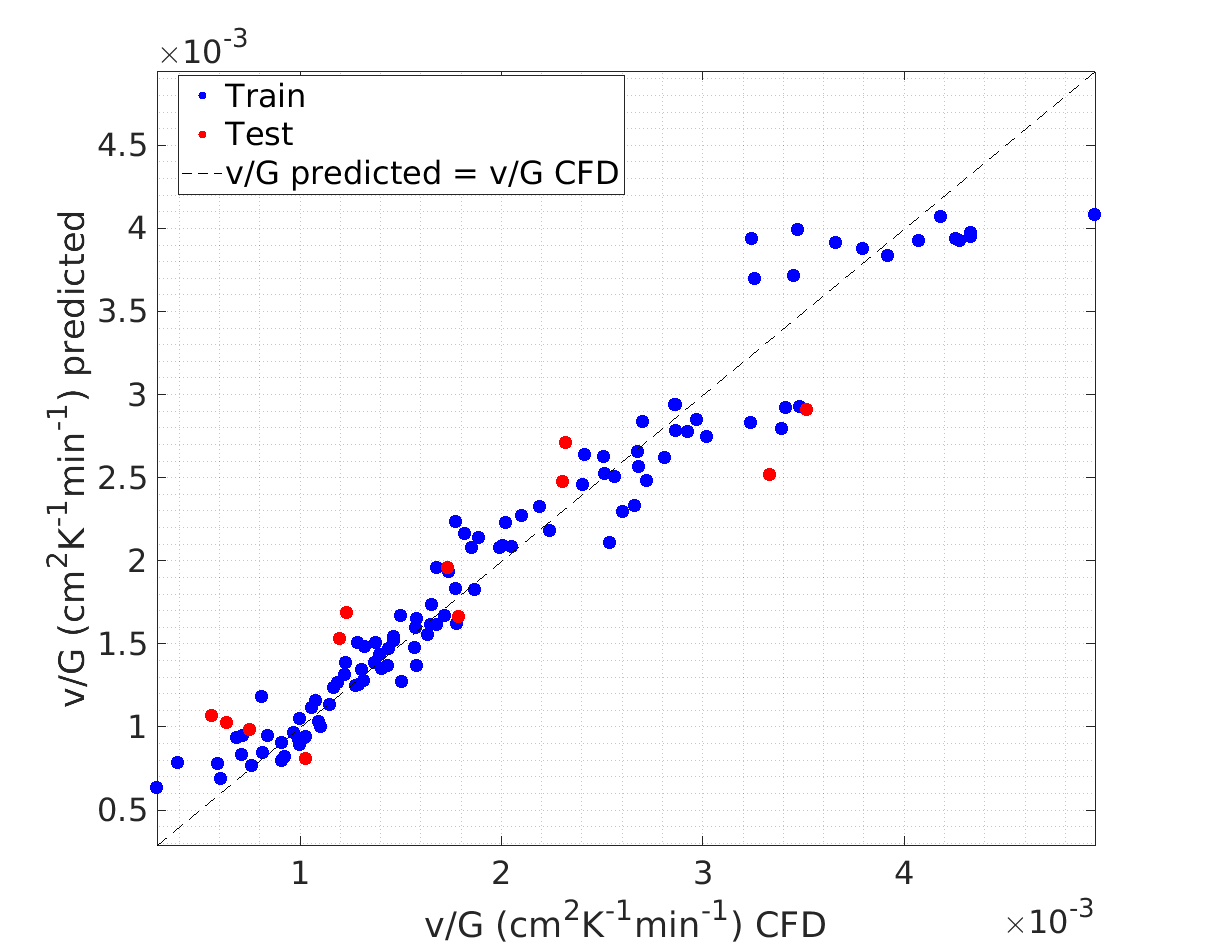}
        \caption{LightGBM model for Ge $V/G$ prediction}
        \label{fig:LGBM_vg_Ge}
    \end{subfigure}
        \begin{subfigure}[b]{0.48\textwidth}
        \centering
    \includegraphics[width=\textwidth]{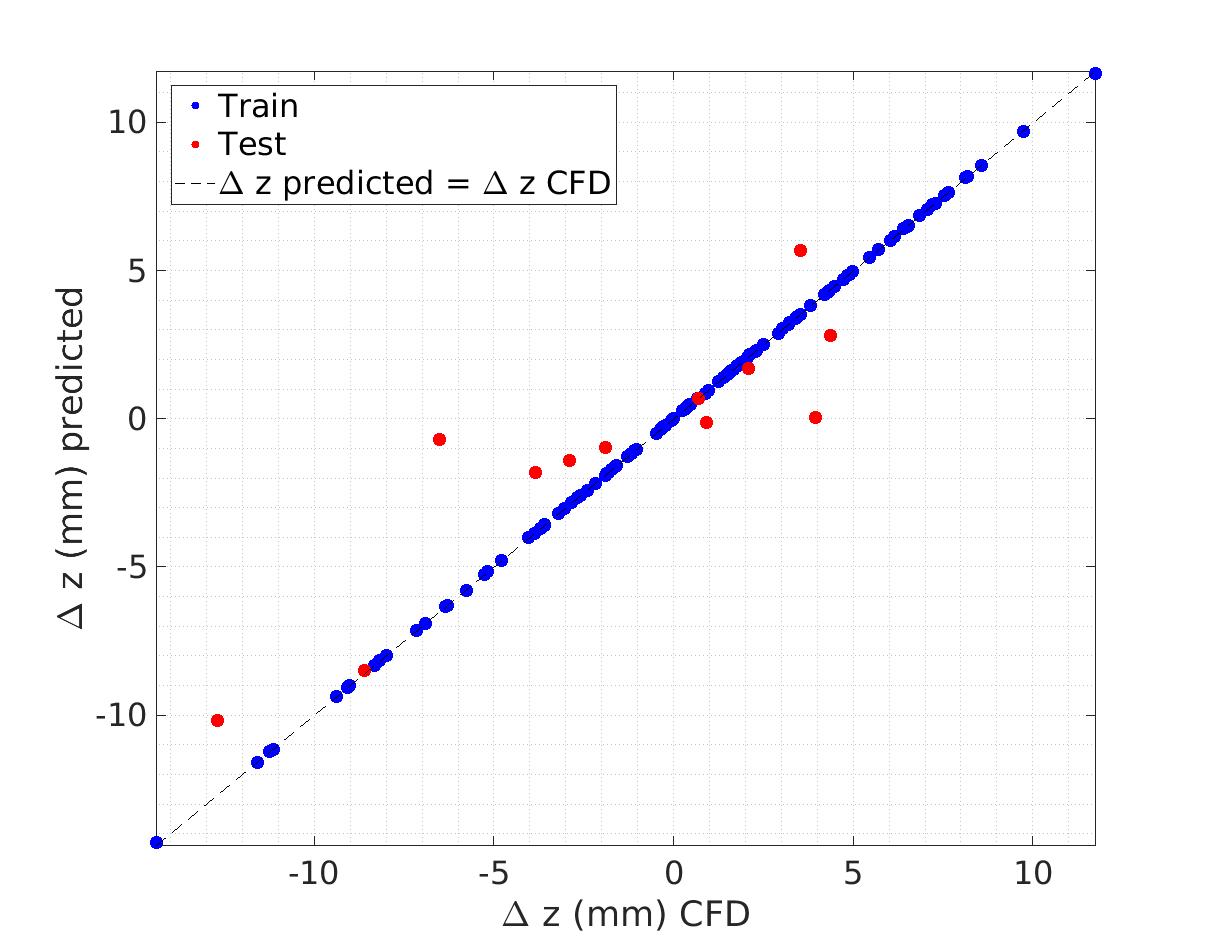}
        \caption{HPT XGBoost model for Ge $\Delta z$ prediction}
        \label{fig:HPT_XGB_Ge}
    \end{subfigure}
    \hfill
    \begin{subfigure}[b]{0.48\textwidth}
        \centering
        \includegraphics[width=\textwidth]{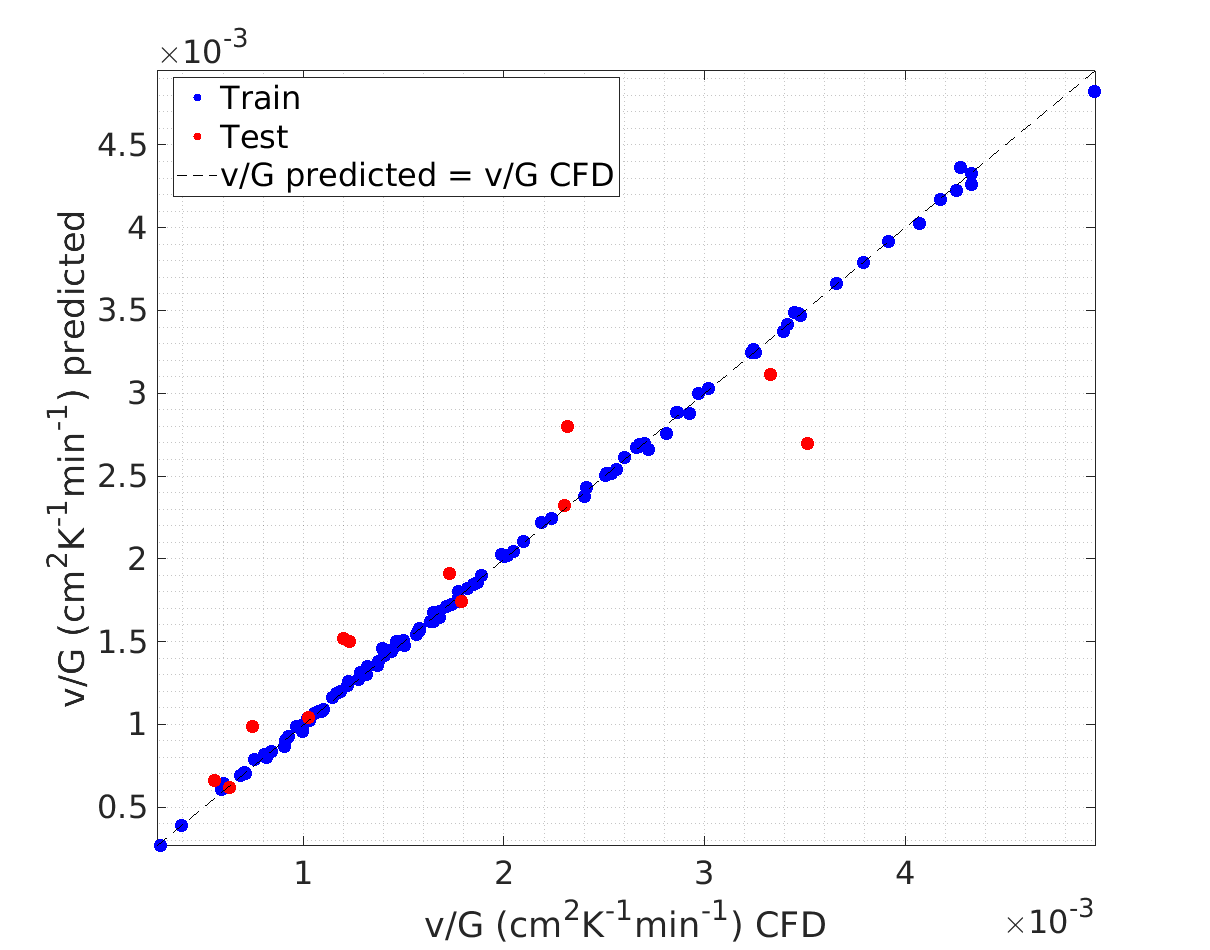}
        \caption{MT LightGMB model for Ge $v/G$ prediction}
        \label{fig:MT_LGBM_Ge}
    \end{subfigure}
    \caption{ Parity plots for Germanium $\Delta z$ and $v/G$ models: \textcolor{blue}{\textbullet} represents training set instances and \textcolor{red}{\textbullet} test set instances }
    \label{fig:parity_plot_ge}
\end{figure}

Furthermore, Figure~\ref{fig:parity_plot_ge} shows parity plots of different models for predicting $\Delta z$ and $v/G$ for Germanium. Figures~\ref{fig:XGB_def_Ge} and~\ref{fig:LGBM_vg_Ge} show the performance of XGBoost and LightGBM models tuned and trained on the Germanium dataset. It can be noticed that for both output parameters, when applying transfer learning, the accuracy increases, and the models generalize better.

\begin{figure}[ht!]
    \centering
    \begin{subfigure}[b]{0.48\textwidth}
        \centering
        \includegraphics[width=\textwidth]{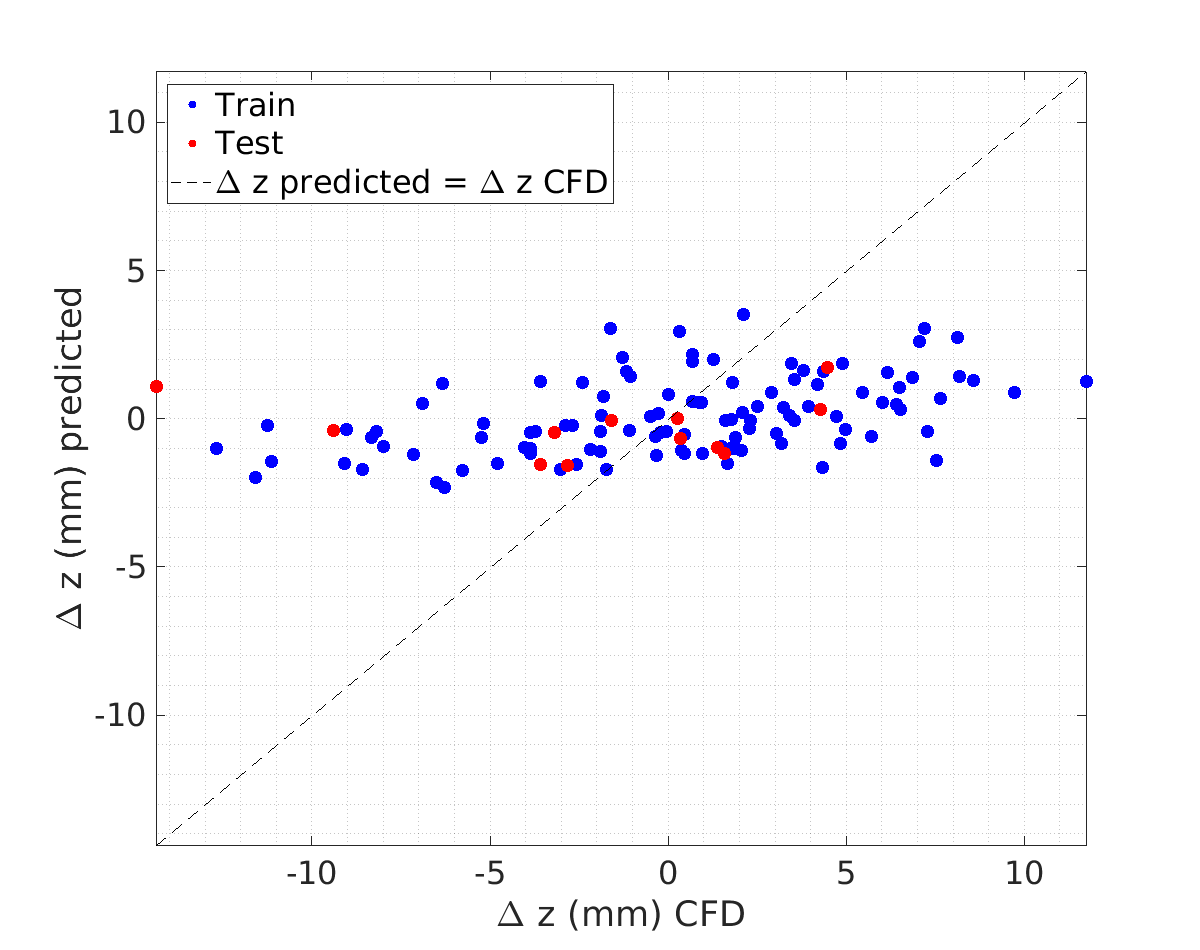}
        \caption{MLP model for GaAs $\Delta z$ prediction}
        \label{fig:MLP_GE}
    \end{subfigure}
    \hfill
    \begin{subfigure}[b]{0.48\textwidth}
        \centering
        \includegraphics[width=\textwidth]{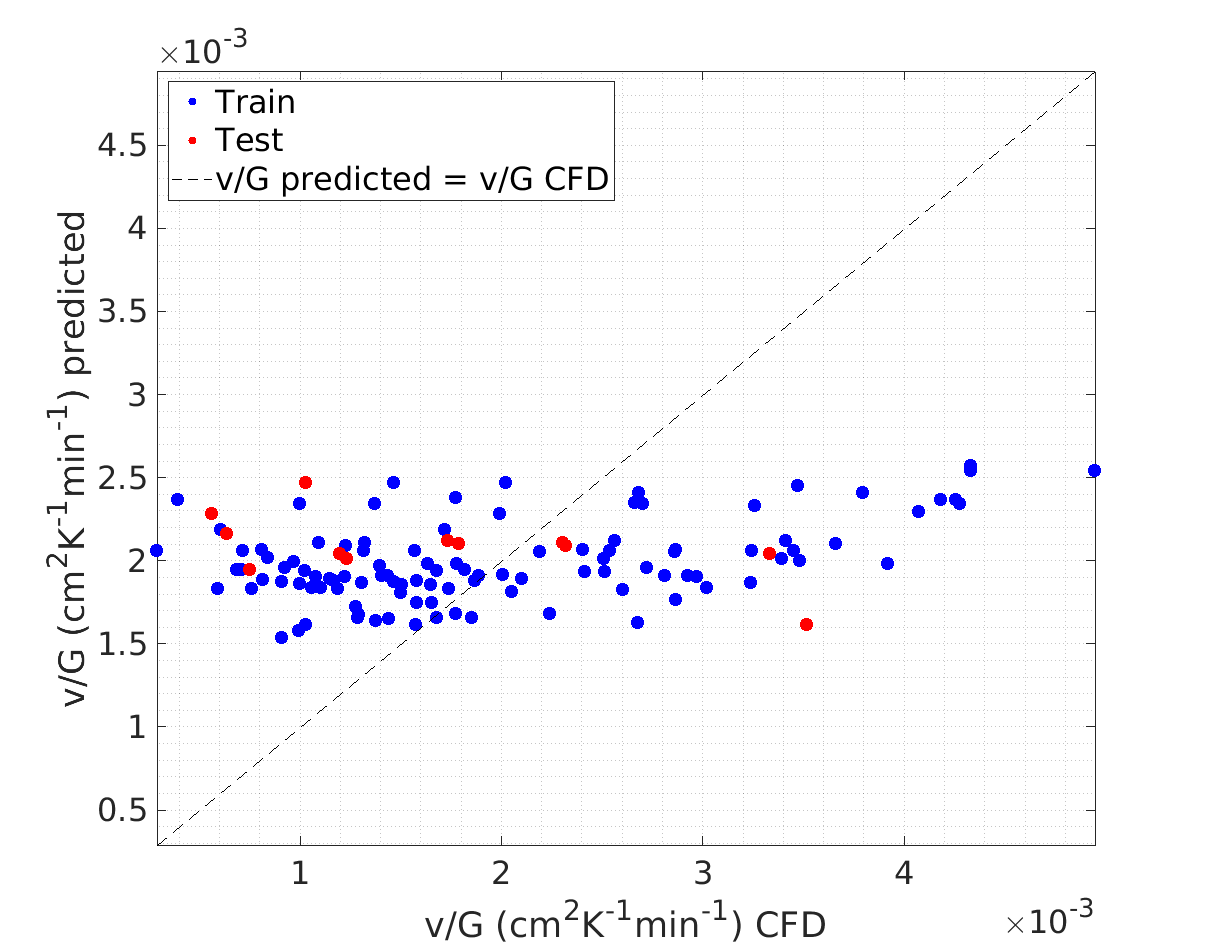}
        \caption{LightGMB model for GaAs $v/G$  prediction}
        \label{fig:LGB_GaAs}
    \end{subfigure}
        \begin{subfigure}[b]{0.48\textwidth}
        \centering
    \includegraphics[width=\textwidth]{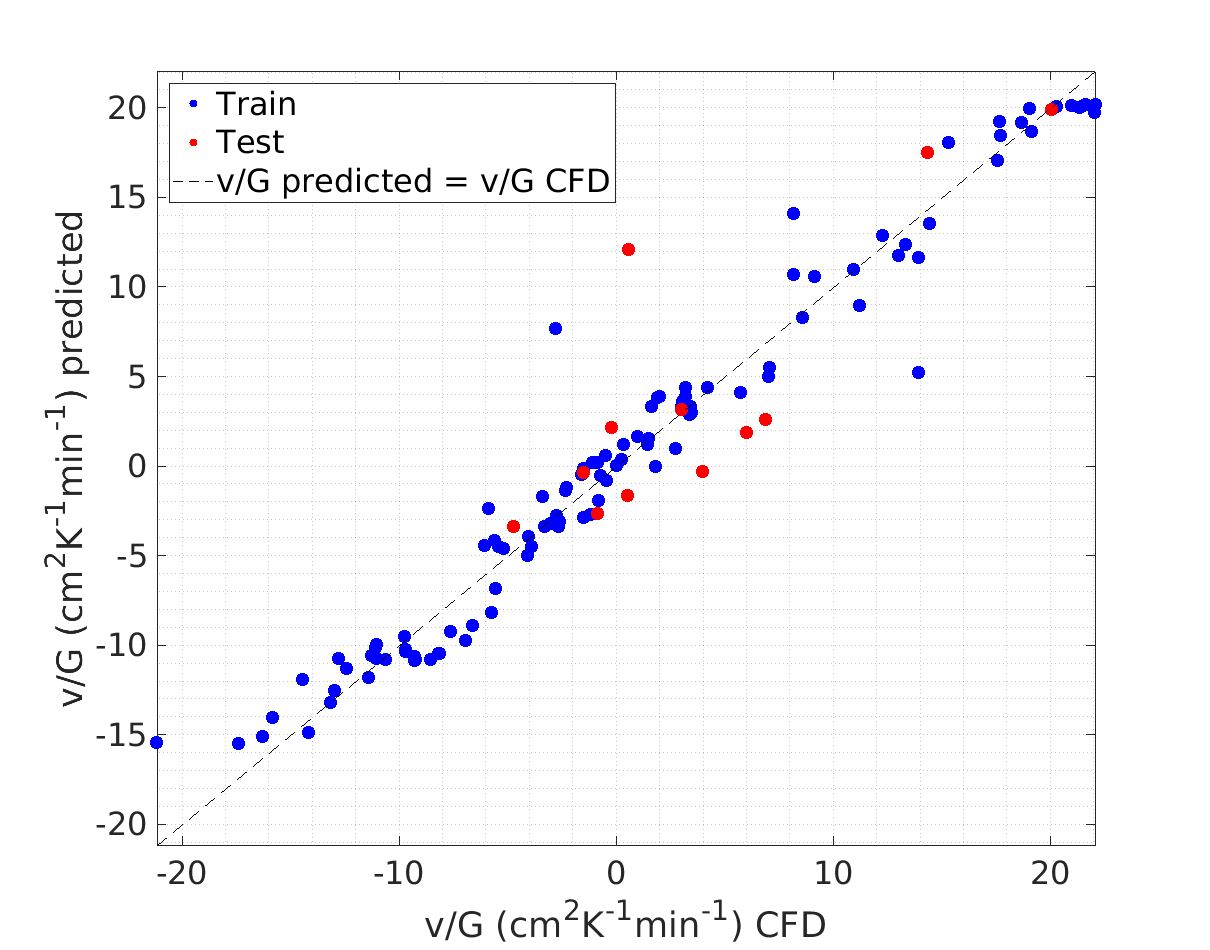}
        \caption{MT LightGMB model for Ge $v/G$  prediction}
        \label{fig:MT_LGB_GE}
        
    \end{subfigure}
    \hfill
    \begin{subfigure}[b]{0.48\textwidth}
        \centering
        \includegraphics[width=\textwidth]{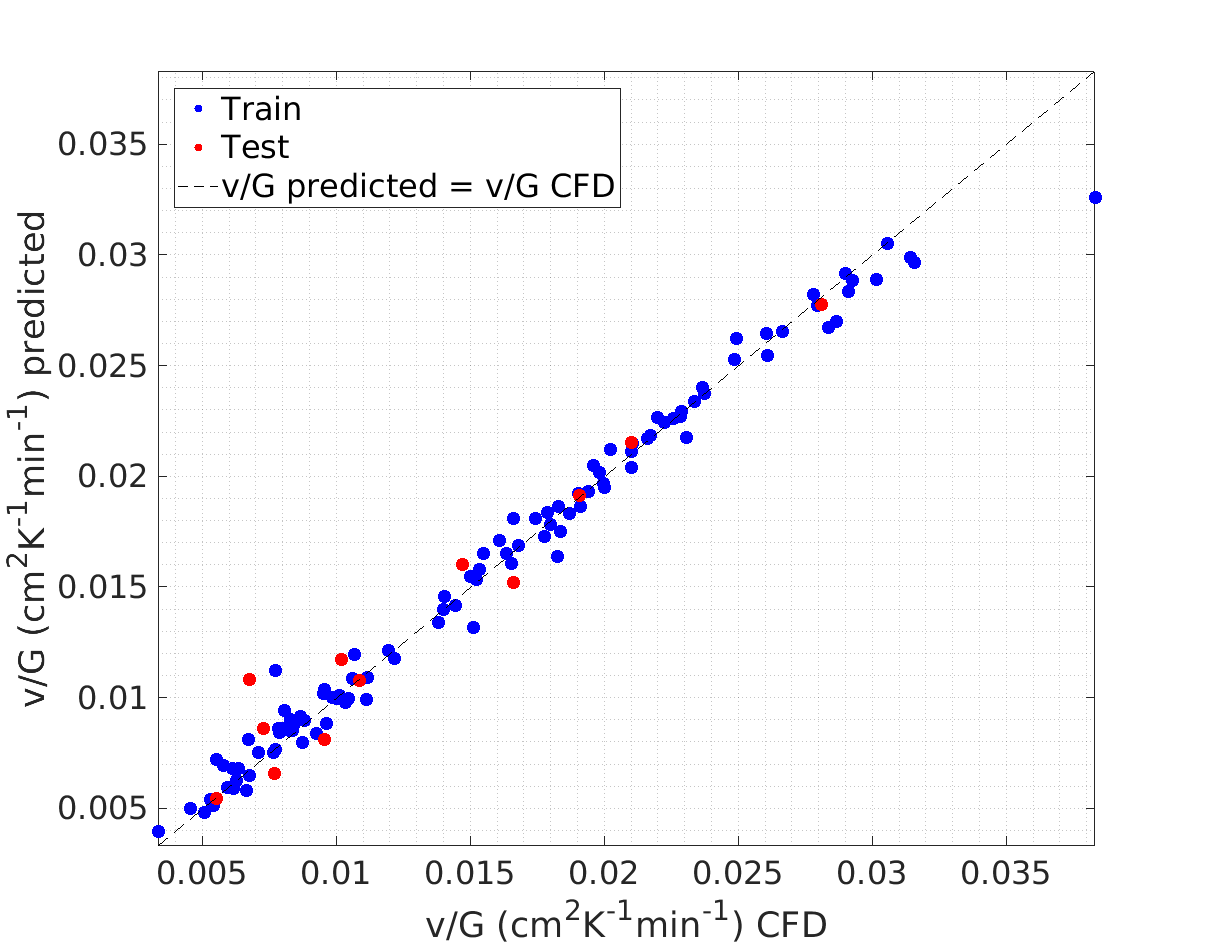}
        \caption{MT LightGMB model for GaAs $v/G$  prediction}
        \label{fig:MT_LGB_GaAs}
    \end{subfigure}
    \caption{ Parity plots for Gallium Arsenide $\Delta z$ and $v/G$ models: \textcolor{blue}{\textbullet} represents training set instances and \textcolor{red}{\textbullet} test set instances }
    \label{fig:parity_plot_vg}
\end{figure}

Similarly, Figure~\ref{fig:parity_plot_vg} shows parity plots for different models predicting $\Delta z$ and $v/G$ for Gallium Arsenide. In this case, both base models trained exclusively on GaAs data fail to capture complex relations between input and output variables. When applying transfer learning methodologies, the accuracy increases significantly on both training and test sets.

XGBoost and LightGBM are robust to small datasets, particularly in the transfer learning context. These models can generalize across materials effectively by adapting to new data with fewer instances. Their decision-tree-based nature allows them to handle non-linear relationships and extract meaningful patterns even from limited data. XGBoost, in particular, shows a stable performance across all materials and transfer techniques, making it the most reliable model for this task.

\section{Conclusion}
\label{sec:conclusion}
This study demonstrates how transfer learning can effectively address high-quality experimental or simulation data scarcity in Czochralski (Cz) crystal growth. By using robust tree-based algorithms (XGBoost and LightGBM) and a feedforward neural network (MLP), we systematically explored three transfer learning strategies (Warm Start, Merged Training, and Hyperparameters Transfer) across three semiconductor materials (Si, Ge, and GaAs). Our findings indicate that tree-based models are particularly well-suited for smaller datasets, outperforming the MLP in baseline and transfer scenarios. XGBoost and LightGMB, in particular, exhibited the highest predictive accuracy and the greatest gains when trained with combined or adapted data from multiple materials.

Moreover, the Hyperparameter Transfer technique proved especially beneficial for bridging materials with differing thermophysical characteristics, leading to notable reductions in prediction error for interface deflection ($\Delta z$) and the Voronkov ratio ($v/G$). In cases where the source and target materials had sizable differences in their underlying feature distributions, even modest amounts of target-domain data (e.g., from GaAs) were enough to boost predictive performance once transfer learning was applied. While MLP remained more data-hungry and sensitive to how parameters were transferred, it benefited substantially from Warm Start when leveraging knowledge gleaned from the more data-rich Si domain.

The results highlight that careful application of transfer learning can significantly reduce modeling errors in bulk crystal growth processes while minimizing the need for extensive new datasets. This streamlines the design and optimization of growth parameters and underscores the potential of hybrid physics-based and data-driven workflows to accelerate materials research. Future studies can extend this framework to additional material systems, incorporate active learning to refine data sampling strategies, and integrate physical constraints into the transfer learning pipeline more tightly. Through these enhancements, data-driven modeling in crystal growth stands to become more versatile and robust, benefiting researchers and industry practitioners alike.

\medskip
%\textbf{Supporting Information} \par %Please delete the Suppporting Information statement if it is not applicable. Please supply Supporting Information in another file. Supporting information should not be provided in .tex format
%Supporting Information is available from the Wiley Online Library or from the author.

% Acknowledgements
\medskip
%\textbf{Acknowledgements} \par %delete if not applicable))
%Please insert your acknowledgements here
\textbf{Acknowledgments} \par %delete if not applicable))
The authors gratefully acknowledge Martin Handwerg for proofreading the manuscript. The research was partly funded by the German Research Foundation, project number 467401796. 
% References
\medskip

% Use the following code if you wish to generate your bibliography with BibTeX;
% replace the string "MSP-template" below with the name(s) of
% the BibTeX data base(s) you want to use.
% The resulting bibliography-output (the content of the .bbl file)
% must be pasted back into this file before submission.
% Please also include your BibTeX data base file(s) in your submission
% so that we can re-run BibTeX if necessary.
%
\bibliographystyle{elsarticle-num} 
\bibliography{references}

\end{document}